%% file: ms.tex
\documentclass{emulateapj-rtx4}
\newcommand{\kt}{\ensuremath{k_{\rm{B}}T}}

\newcommand{\fx}{\ensuremath{F_{\rm{X}}}}
\newcommand{\nh}{\ensuremath{N_{\rm H}}}

\newcommand{\ergcms}{{\rm ergs}\ {\rm cm}^{-2}\ {\rm s}^{-1}}

\newcommand{\chandra}{\textit{Chandra}}
\usepackage{natbib}

\slugcomment{Accepted to \apj\ (Version January 25, 2013)}

\shorttitle{X-ray Point Source Populations for GRXE}
\shortauthors{Morihana et al.}

\begin{document}
\title{X-ray Point Source Populations Constituting the Galactic Ridge X-ray Emission}
\author{Kumiko Morihana\altaffilmark{1}, Masahiro Tsujimoto\altaffilmark{2}, Tessei
Yoshida\altaffilmark{3}, and Ken Ebisawa\altaffilmark{2,4}
}
\email{morihana@crab.riken.jp}
\altaffiltext{1}{Institute of Physical and Chemical Research (RIKEN), 2-1 Hirosawa, Wako, Saitama 351-0198, Japan}
\altaffiltext{2}{Japan Aerospace Exploration Agency, Institute of Space and Astronautical Science, 3-1-1 Yoshino-dai, Chuo-ku, Sagamihara, Kanagawa 252-5210, Japan}
\altaffiltext{3}{National Astronomical Observatory of Japan, 2-21-1, Osawa, Mitaka, Tokyo 181-8588, Japan}
\altaffiltext{4}{Department of Astronomy, Graduate School of Science, The University of Tokyo, 7-3-1 Hongo, Bunkyo-ku, Tokyo 113-0033, Japan}

\begin{abstract}
 Apparently diffuse X-ray emission has been known to exist along the central quarter of
 the Galactic Plane since the beginning of the X-ray astronomy, which is referred to as
 the Galactic Ridge X-ray emission (GRXE). Recent deep X-ray observations have shown
 that numerous X-ray point sources account for a large fraction of the GRXE in the hard
 band (2--8~keV). However, the nature of these sources is poorly understood. Using the
 deepest X-ray observations made in the \textit{Chandra} Bulge Field
 \citep{revnivtsev09,revnivtsev11}, we present the result of a coherent photometric and
 spectroscopic analysis of individual X-ray point sources for the purpose of
 constraining their nature and deriving their fractional contributions to the hard band
 continuum and Fe K line emission of the GRXE. Based on the X-ray color-color diagram,
 we divided the point sources into three groups: A (hard), B (soft and broad spectrum),
 and C (soft and peaked spectrum). The group A sources are further decomposed spectrally
 into thermal and non-thermal sources with different fractions in different flux
 ranges. From their X-ray properties, we speculate that the group A non-thermal sources
 are mostly AGNs and the thermal sources are mostly white dwarf (WD) binaries such as
 magnetic and non-magnetic cataclysmic variables (CVs), pre-CVs, and symbiotic stars,
 whereas the group B and C sources are X-ray active stars in flares and quiescence,
 respectively. In the $\log N$--$\log S$ curve of the 2--8~keV band, the group A
 non-thermal sources is dominant above $\approx$10$^{-14}$~$\ergcms$, which is gradually
 taken over by Galactic sources in the fainter flux ranges. The Fe K$\alpha$ emission is
 mostly from the group A thermal (WD binaries) and the group B (X-ray active stars)
 sources.
\end{abstract}


\section{Introduction}\label{s1}
Since the dawn of the X-ray astronomy, the apparently diffuse emission of low surface brightness has been known to exist along the Galactic Plane (GP; $|l|< $45$^{\circ}$, $|b|<
$1$^{\circ}$), which is referred to as the Galactic Ridge X-ray emission (GRXE;
e.g.,~\citealt{worrall82};~\citealt{warwick85};~\citealt{koyama86}).  The X-ray spectrum
is characterized by hard continuum with a strong Fe K emission feature at 6--7~keV band
\citep{koyama86}. The GRXE is considered to have different origins for the soft
($\lesssim$2~keV) and hard ($\gtrsim$2~keV) emission (e.g., \citealt{hands04}), and we
focus on the latter in this paper.

Recently, \citet{revnivtsev09} has shown that the majority ($\sim$80\%) of the Fe K-band
emission was resolved into point sources using the deepest X-ray observations made
with the \textit{Chandra X-ray Observatory} \citep{weisskopf02} at a slightly off-plane
region of ($l$, $b$)$=$(0\fdg113, --1\fdg424) in the Galactic bulge (\textit{Chandra}
bulge field; CBF). The sensitivity of the study reached $\sim$10$^{-16}$ $\ergcms$ in
the hard (2--8~keV) X-ray band with a total integration time of nearly 1~Ms, which is
$\gtrsim$10 times deeper than all previous studies. In previous shallower surveys, only
a small fraction of the GRXE was resolved into point sources. For example,
\citet{ebisawa01, ebisawa05} conducted 100~ks \textit{Chandra} observations at two partially
overlapping regions in the GP at ($l$, $b$)$=$(28\fdg5, 0\fdg0), and found that only
$\sim$10\% of the total GRXE emission was resolved into point sources. These
findings suggest that a large contribution is made for the GRXE by point sources in
the flux range around 10$^{-15}$--10$^{-16}$~$\ergcms$.

The nature of these numerous dim point sources remains unknown. They may be explained by
the extrapolation of the X-ray source population known in brighter flux ranges, or new
classes of X-ray sources are required as major contributers at this flux range
\citep{ebisawa05}. Understanding the X-ray source population at a certain flux range
boils down to constructing the $\log{N}-\log{S}$ curve separately for major classes of
sources. In the hard X-ray band, the $\log{N}-\log{S}$ curve was studied with the
\textit{Advanced Satellite for Cosmology and Astrophysics} \citep{sugizaki01} at a flux
range brighter than $\sim$10$^{-12.5}$~$\ergcms$ and with the \textit{Chandra} and
\textit{XMM-Newton} Observatories \citep{ebisawa05,hands04,motch10} at a flux range
between $\sim$10$^{-12.5}$ and $\sim$10$^{-14.5}$~$\ergcms$. These studies divided
detected point sources phenomenologically based on the X-ray properties aided by optical
and near-infrared identifications, and revealed that several classes of sources
contribute to the $\log{N}-\log{S}$ curve with different fractions in different flux
ranges. The brightest end of the curve is dominated by low-mass X-ray binaries, which
saturates at $\sim$10$^{-13}$~$\ergcms$ for being integrated all throughout our Galaxy
in the line of sight. Cataclysmic variables (CVs) and X-ray active binary stars emerge
as dominant contributers below the flux. Extra-galactic sources, which are mostly active
galactic nuclei (AGNs), co-exist with these Galactic sources in survey fields in all
flux ranges.

Even fainter flux range is accessible only with the deepest \textit{Chandra} observation
in the CBF. \citet{revnivtsev11} constructed the $\log{N}-\log{S}$ curve with all point
sources in the central 2$\farcm$56 circle within 17\farcm4$\times$17\farcm4 fields (only
7\% in the area) down to 10$^{-16}$~$\ergcms$ and discussed that the curve can be
reproduced by the sum of CVs and active binary stars using the expected number of these
sources in the line of sight and their luminosity function derived in the Solar vicinity
\citep{Sazonov06}. However, in the previous studies in the CBF
\citep{revnivtsev09,revnivtsev11}, individual sources were treated collectively and
their different X-ray properties were not considered, which should be useful to classify
these sources.

What we aim to do in this paper is to derive the X-ray photometric properties (flux and
color) from all the sources in the entire CBF field and the spectroscopic and timing
properties from bright sources through coherent data analysis. By using the entire
field, we can increase the number of sources at a compensation of slightly lower
averaged sensitivity in comparison to the previous studies only using the central part
\citep{revnivtsev09,revnivtsev11}. Then, we divide the sources phenomenologically into
several groups based on these X-ray properties to construct the $\log{N}-\log{S}$ curve
separately for each group to find their contributions at the flux range of interest. We
also consider the contribution of these groups to the GRXE separately for the Fe K
emission and the continuum emission in the hard band, as different classes are expected
to have different equivalent width (EW) of the feature.

\medskip

The outline of this paper is as follows. In $\S~\ref{s2}$, we present observations and
data reduction.  In $\S~\ref{s3}$, we extract point sources, derive the survey
completeness limit, and obtain X-ray properties of individual sources. In $\S~\ref{s4}$,
we divide all the sources based on their X-ray colors and flux, and discuss their likely
classes based on the spectroscopic and timing properties of the brightest members of
each group, and the composite spectrum of all sources in the group. We derive the
contribution of these sources to the GRXE. We summarize the results in
$\S~\ref{s5}$. For clarity, we use ``groups'' for subsets of detected point sources
divided phenomenologically based on their X-ray properties, and ''classes'' for
different types of astronomical objects (e.g., AGNs, CVs, stars, etc) in this paper.

\section{Observations and Data Reduction}\label{s2}
We retrieved ten archived data of the CBF taken with the Advanced CCD Imaging
Spectrometer (ACIS; \citealt{garmire03})-I array on board \chandra. The observations
were carried out from 2008 May to August with a total exposure time
of $\sim$900~ks (Table~\ref{t01}). The ACIS-I covers the 0.5--8.0~keV energy band with a
spectral resolution of $\sim$280~eV for the full width at a half maximum at 5.9~keV as
of the observation dates\footnote{See http://cxc.harvard.edu/cal/Acis/ for details.}.
The CCDs were operated with a frame time of 3.2~s. The data were down-linked with the
very faint telemetry format.

\input{t01}

\begin{figure}[htbp]
 \epsscale{0.6}
 \plotone{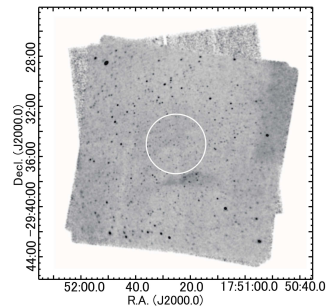}
 \caption{Smoothed and exposure-corrected X-ray image (0.5--8.0~keV). All the ten data
 sets were stacked. The white circle at the center shows the region studied by
 \citet{revnivtsev09,revnivtsev11}.}
 \label{f01}
\end{figure}

We retrieved pipeline products and reduced the data sets using the \chandra~Interactive
Analysis of Observations (CIAO) software package \citep{fruscione06} version~4.2. We
reprocessed the raw data to obtain an X-ray event list for each ObsID, in which we
filtered events based on event grades, good time intervals, and energy bands
(0.5--8.0~keV), and applied a background rejection algorithm to data taken with the very
faint telemetry format\footnote{See
http://cxc.harvard.edu/cal/Acis/Cal\_prods/vfbkgrnd/index.html for detail.}. We then
merged the ten event files into one. Figure~\ref{f01} shows the combined ACIS-I images
corrected for the non-uniformity of the exposure across the study field.

\section{Analysis}\label{s3}
\subsection{Source Extraction}\label{s3-1}
We first extracted point source candidates using the \texttt{wavdetect} algorithm in the
CIAO package. We set the significance threshold at 2.5 $\times$ 10$^{-5}$, implying that
one false positive detection would be expected at every 4 $\times$ 10$^{4}$ trials. As a
result, 2,596 source candidates were found. The number of our source candidates is
nearly the same with that by \cite{revnivtsev09} in the same region.

For all the source candidates, we extracted source and background events using the ACIS
Extract (AE; \citealt{broos10}) package version~2009-12-01. The source events were
extracted from a region containing $\sim$90\% of photons based on the simulated point
spread function of each source, while the background events were extracted locally from
an annulus around each source. When two source extraction regions overlap with each
other, each region was shrunk so that the overlap disappeared. In the background
regions, circles with a radius of 1.1 times the $\sim$99\% encircled energy radius of
each source candidate were masked.  All these processes are automated in the AE package,
which is useful to extract sources in the studied region with a high source surface
number density possibly with underlying extended emission.

\input{t02}

To select significant point sources from the candidates, we examined their validity
based on their photometric significance (PS) and the probability of no source
($P_{\rm{B}}$). The PS is defined as the background-subtracted source counts
($C_{\mathrm{net}}$) divided by its background counts normalized by the area.
$P_{\rm{B}}$ is the probability that the source is attributable to a background
fluctuation assuming the Poisson statistics. We recognized the source to be valid if
they satisfy both two criteria: $\mathrm{PS}$~$\ge$~1.0 and
$P_{\rm{B}}$~$\le$~$1.0\times10^{-2}$. As a result, we obtained 2,002 valid point
sources. Table~\ref{t02} tabulates the photometric properties of the first 20 sources,
and the full source list is available in the on-line version. The X-ray sources can be
referred following the International Astronomical Union (IAU) convention; e.g., CXOU
J175044.88--292837.6 for the source sequence number 1 in Table~\ref{t02}.

\subsection{Detection Completeness}\label{s3-2}
\begin{figure}[hbtp]
 \begin{center}
  \epsscale{0.6}
  \plotone{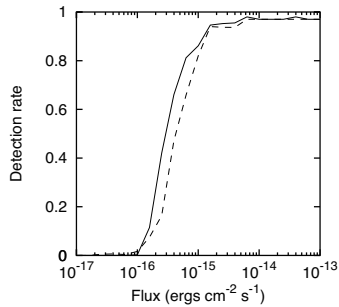}
 \end{center}
 \caption{Detection rate as a function of flux in 0.5--8.0~keV (solid) and in 2--8~keV (dashed).}
\label{f02}
\end{figure}

We estimated the detection completeness averaged over the entire area in the following
method. We generated events of 400 artificial sources using the
\texttt{marx}\footnote{See http://space.mit.edu/CXC/MARX/ for details.} event simulator
at random positions across the image with a given flux. The events were merged with the
observed events, from which point sources were detected using the same algorithm with
the observed data. We derived the detection rate at each flux, which was changed from
10$^{-13}$ to 10$^{-17}~\ergcms$ in the total (0.5--8.0~keV) and the hard (2--8~keV)
bands. We assumed that the artificial sources have the same spectrum with the best-fit
model describing the stacked spectrum of all point sources
(\S~\ref{s4-2}). Figure~\ref{f02} shows the detection rate as a function of flux. The
rate is nearly complete (more than 97\%) above $\sim$10$^{-15.2}~\ergcms$ in the total
band and $\sim$10$^{-14.8}~\ergcms$ in the hard band.

\subsection{Photometry}\label{s3-3}
\begin{figure}[hbtp]
 \begin{center}
 \epsscale{1.0}
  \plotone{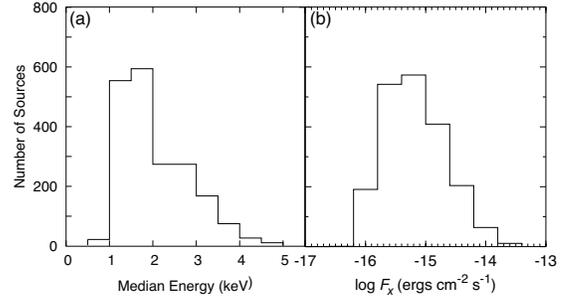}
 \end{center}
 \caption{Distribution of (a) median energy and (b) photometric flux in the 0.5--8.0~keV
 band for all sources.}
\label{f03}
\end{figure}

Because most of the detected sources are faint, we used the photometric information to
characterize their spectral hardness (color) and flux based on a quantile method
\citep{hong04}. We derived the median energy (ME) for each source, which is a proxy for the
spectral hardness. The distribution of the ME of all sources is shown in
Figure~\ref{f03}~(a).

\begin{figure}[htbp]
\epsscale{0.6}
 \plotone{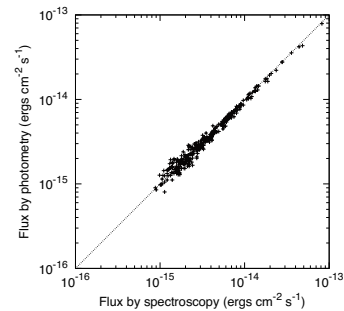}
 \caption{Relation between two X-ray flux estimates for 335 sources, for which spectral
 fitting results are available. The vertical and horizontal axes represent the X-ray
 flux in the 0.5--8.0~keV estimated by the photometric and spectroscopic method,
 respectively. The dotted line represents the equal value between the two estimates.}
 \label{f04}
\end{figure}

We derived the flux of all sources using the photometric flux ($F_{\mathrm{ph}}$)
defined as 
\begin{eqnarray}
 F_{\mathrm{ph}}~(\ergcms) = 1.6\times10^{-9}~\frac{\mathrm{ME~(keV)}~C_{\mathrm{net}}~(\mathrm{s}^{-1})}{\mathrm{EA~(\mathrm{cm}^{2})}}~\nonumber,
\end{eqnarray}
in which EA is the energy-averaged effective area for each source. We derived both the
total (0.5--8.0~keV) and the hard (2--8~keV) band photometric flux. Some sources have no
hard-band photometric flux because of their lack of photons in the band.  The
distribution of $F_{\mathrm{ph}}$ is shown in Figure~\ref{f03}~(b). In order to check
the consistency with the flux derived by spectral fitting (\S~\ref{s3-5}), we compared
the flux estimates both by photometric and spectroscopic methods for 335 sources, for
which spectral fitting result was available. The comparison shows that they are in a
good agreement (Fig.~\ref{f04}). Hereafter, we use the photometric flux for all sources.

\subsection{Variability}\label{s3-4}
In order to examine time variability in the count rate, we applied the
Kolmogorov-Smirnov (KS) test for the sources with net counts more than 100 (355
sources in total), which we call \textit{bright} sources. The test examines the degree
of non-uniformity in the distribution of photon arrival times against a uniform
distribution (or constant flux) using the $\chi^{2}$ statistics. Sources were removed if
they were in the chip gaps or at the field edge in at least one observation, because the
telescope dithering causes artificial variability.

The null hypothesis probability of the KS test was derived for each source in each of the
ten observations. We defined the merged probability ($P_{\mathrm{KS}}$) for each source
as the smallest value among the ten probabilities. Based on this, we divided sources
into three sets: (a) $P_{\mathrm{KS}}\ge 5 \times 10^{-2}$ for non-variable sources
(109 sources), (b) $5\times 10^{-3} \le P_{\mathrm{KS}} < 5 \times10^{-2}$ for
marginally variable sources (68 sources), and (c) $P_{\mathrm{KS}} <5\times10^{-3}$ for
variable sources (34 sources).

\begin{figure*}[htbp]
 \includegraphics[angle=90,width=0.33\textwidth,clip]{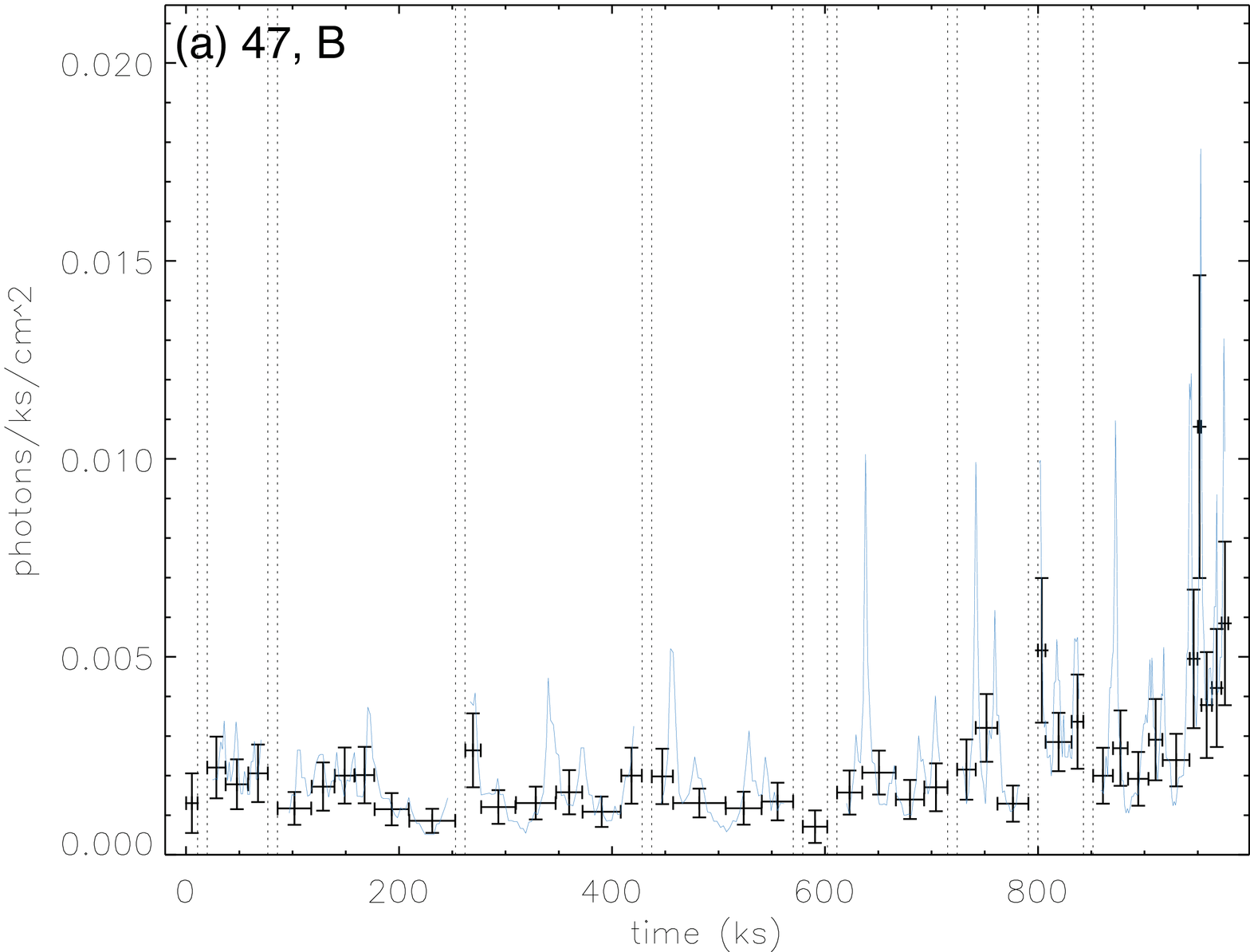}
 \includegraphics[angle=90,width=0.33\textwidth,clip]{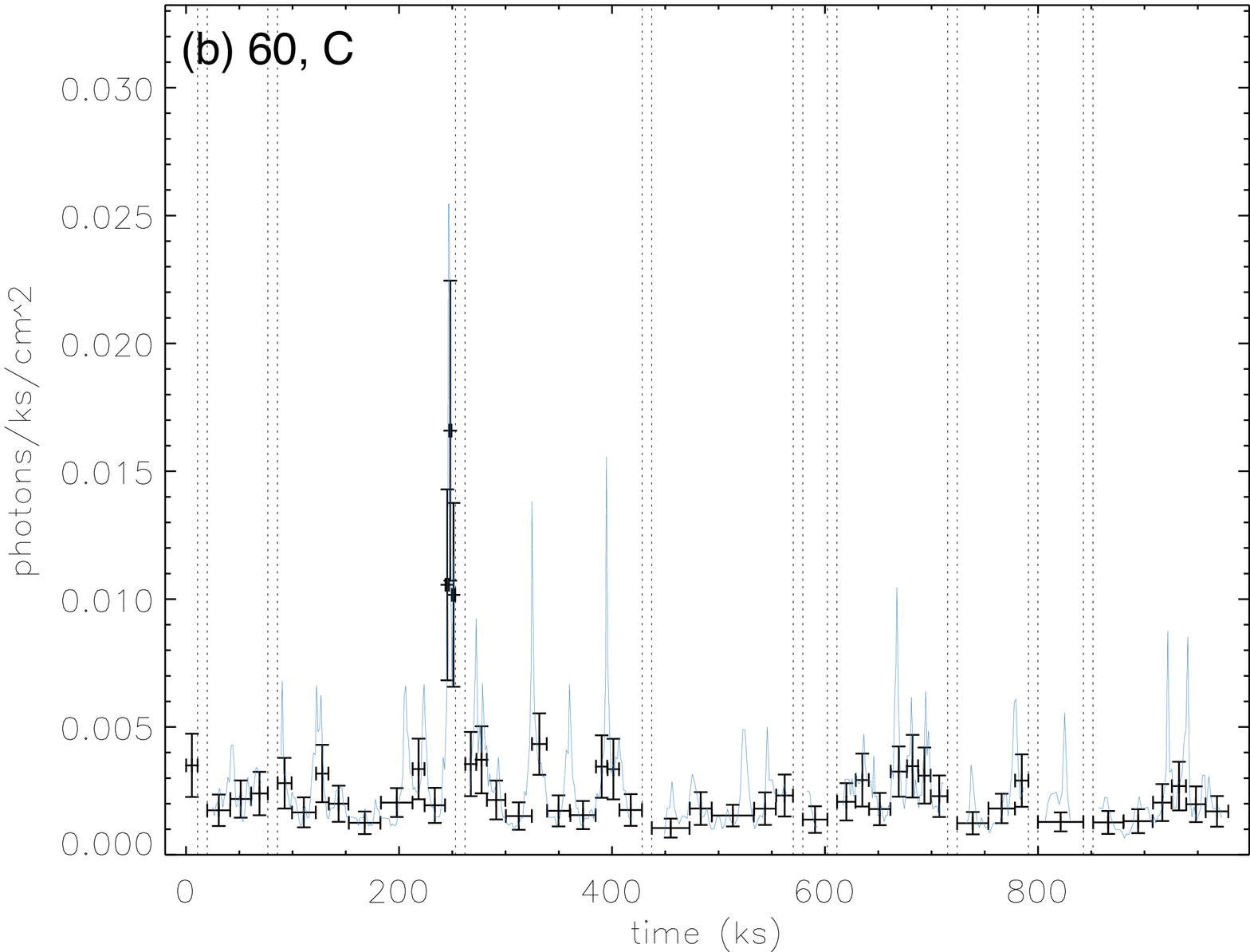}
 \includegraphics[angle=90,width=0.33\textwidth,clip]{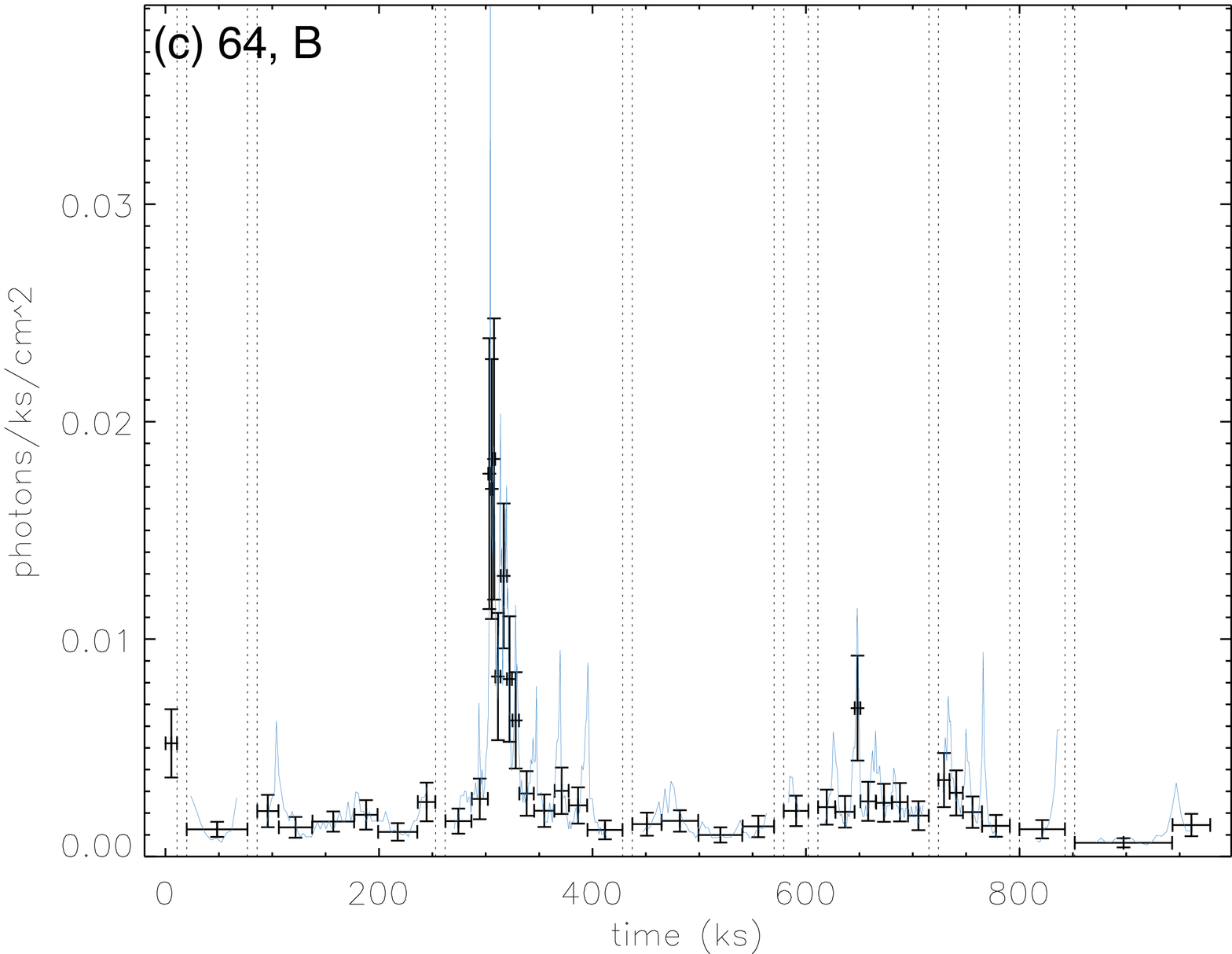}
 \includegraphics[angle=90,width=0.33\textwidth,clip]{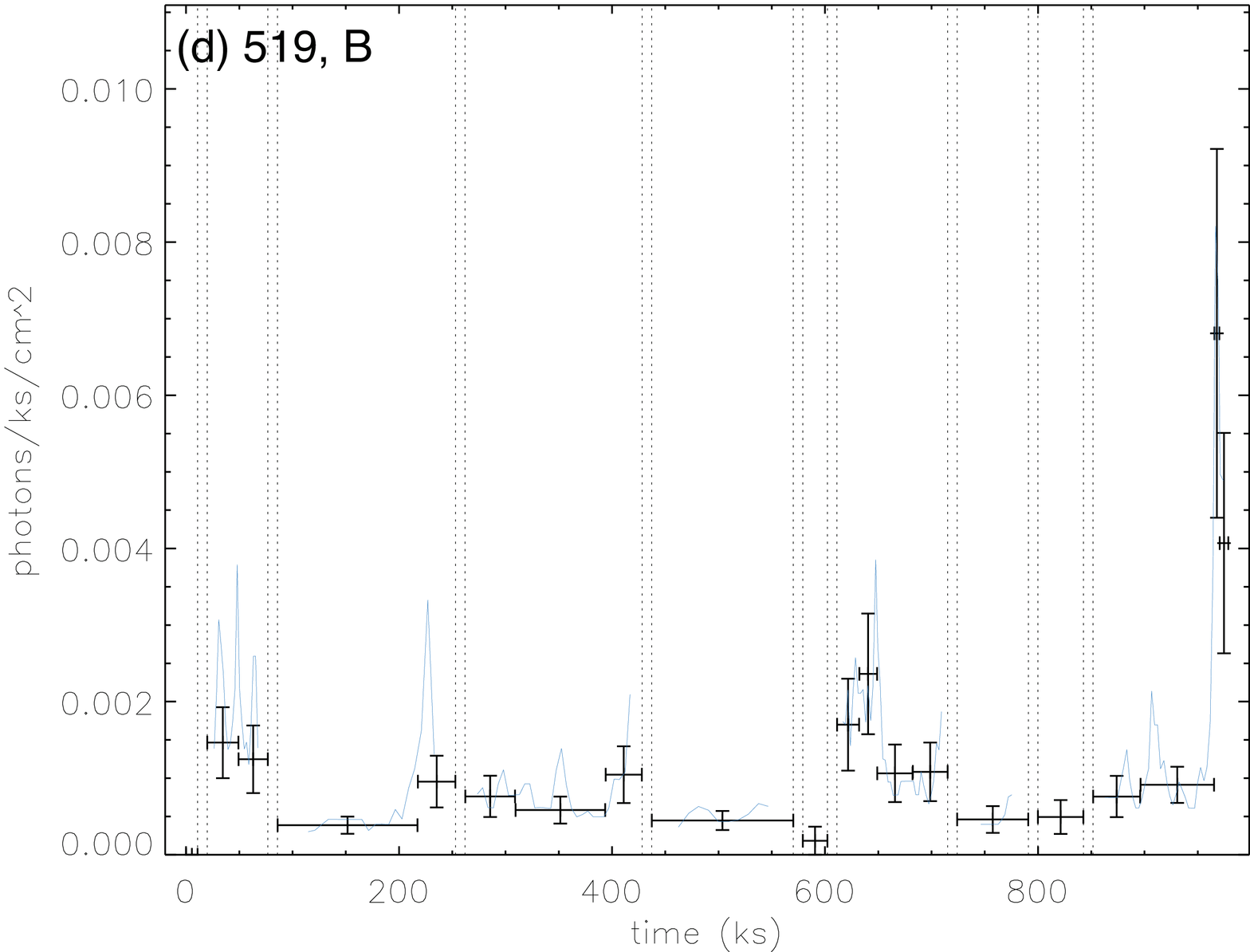}
 \includegraphics[angle=90,width=0.33\textwidth,clip]{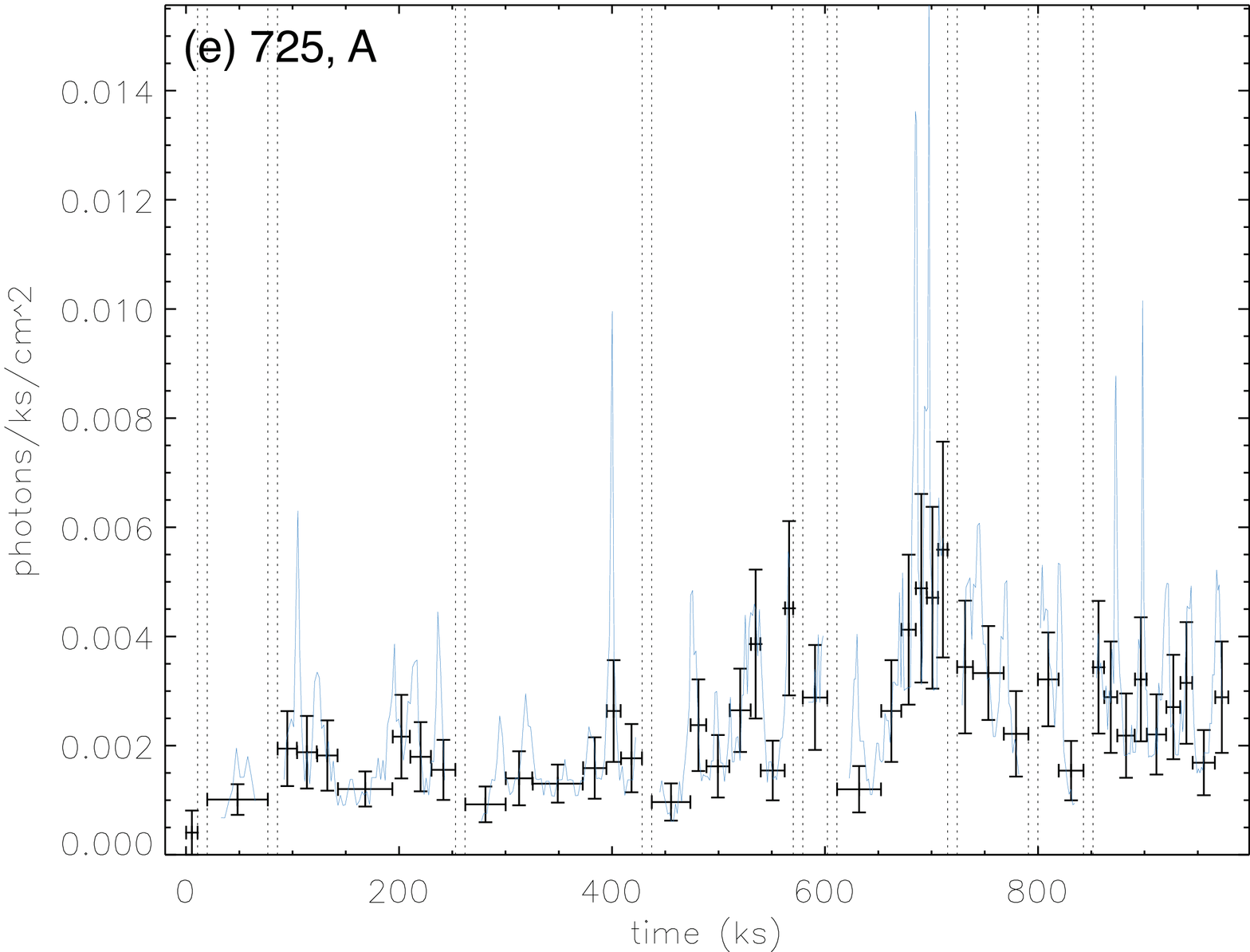}
 \includegraphics[angle=90,width=0.33\textwidth,clip]{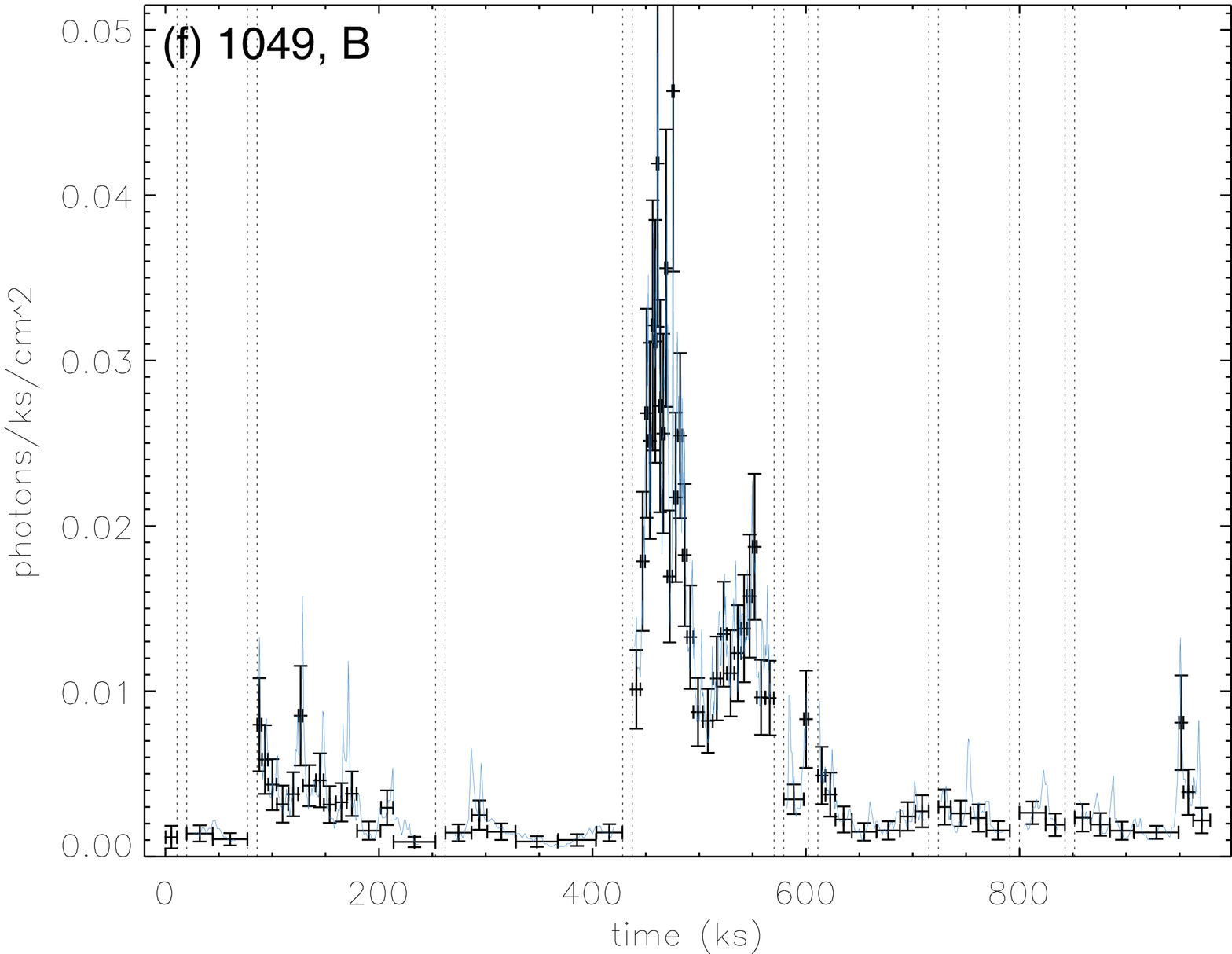}
 \includegraphics[angle=90,width=0.33\textwidth,clip]{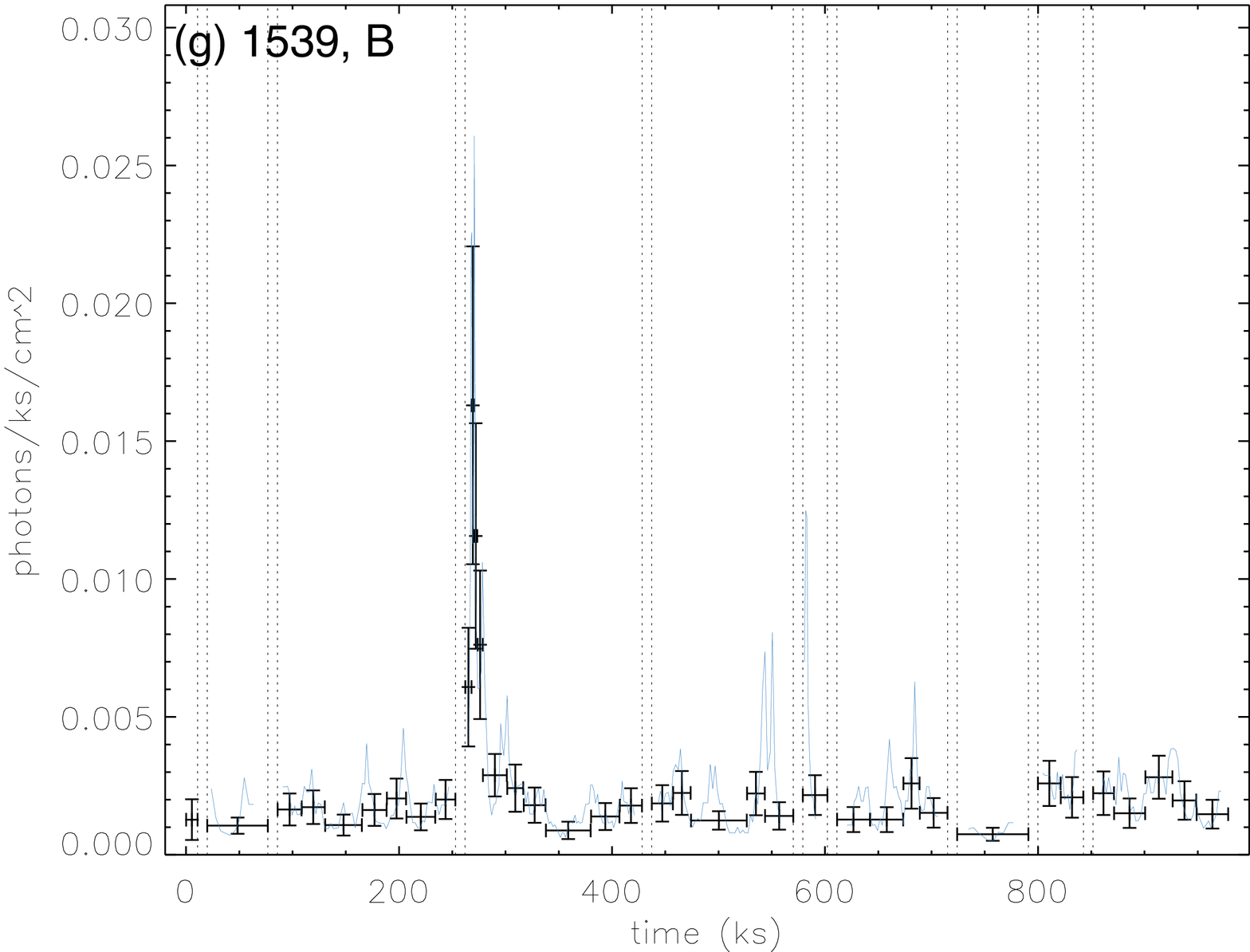}
 \includegraphics[angle=90,width=0.33\textwidth,clip]{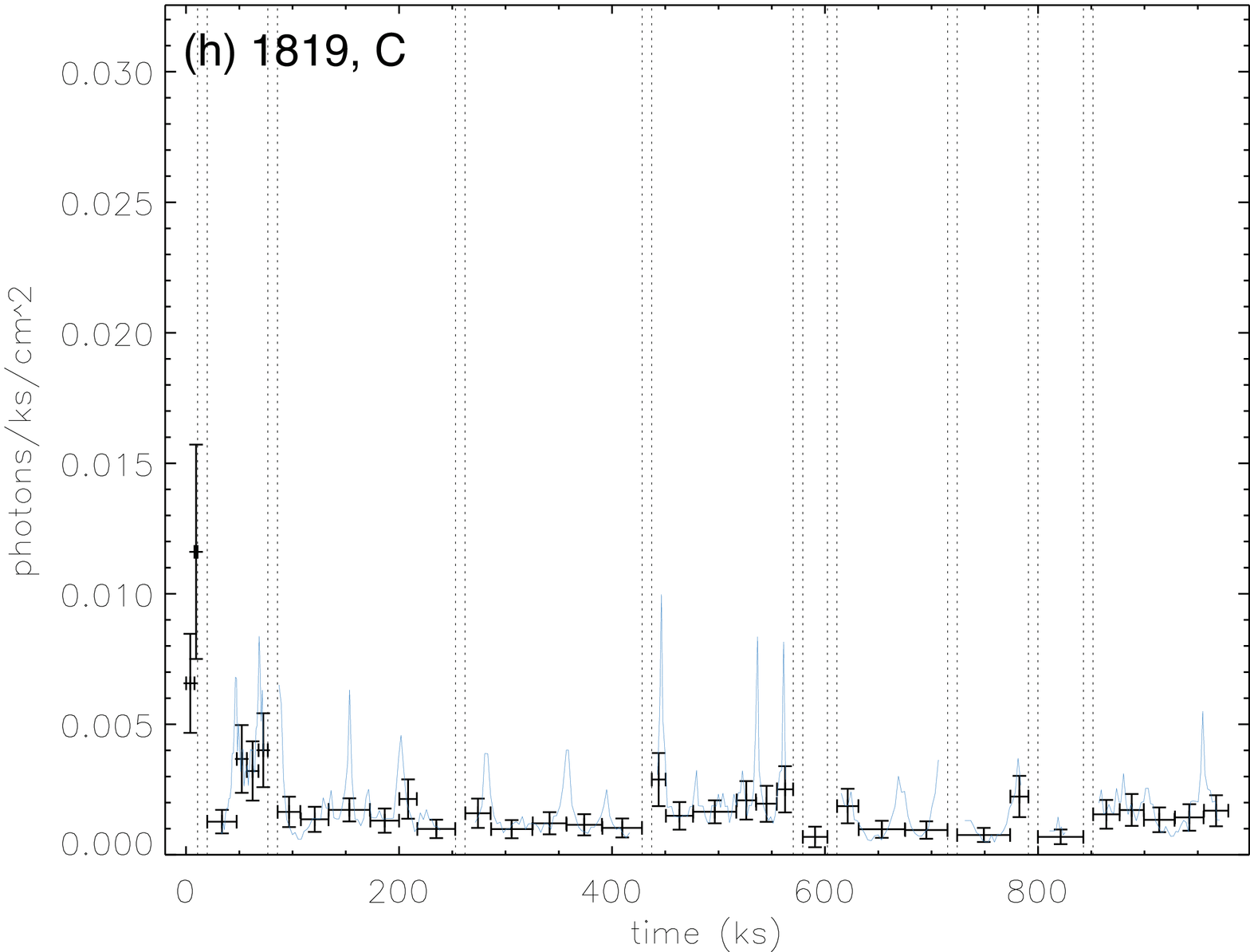}
 \includegraphics[angle=90,width=0.33\textwidth,clip]{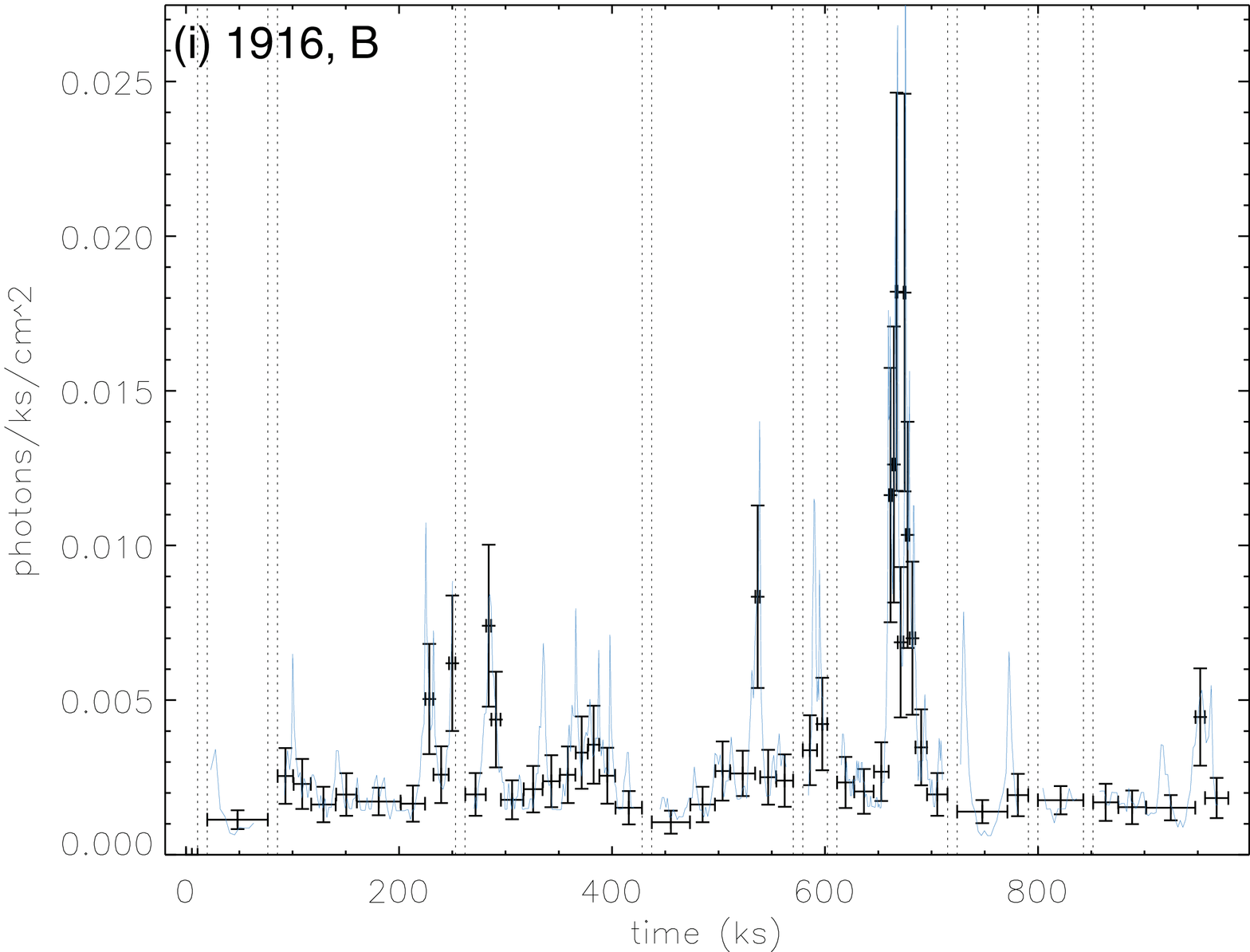}
 \caption{Concatenated light curves for sources with flare-like variability. The light
 curves show variations in count rate (black histograms are binned and blue curves are
 adaptively smoothed). Time intervals between observations are indicated by black dotted
 vertical lines. The top left labels show the sequence number in Table~\ref{t02} and the
 group (\S~\ref{s4-1}) of each source.}
 \label{f05}
\end{figure*}

By looking at the light curve of variable sources, we noticed that some sources show a
typical flare-like flux variation of fast rise and slow decay. The KS test does not
consider the profile of the variation in the light curve, so we picked up flare-like
sources by setting a different criterion that the contrast of the maximum to minimum
count rate is larger than 10 in at least one observation. In this manner, we selected
nine flare-like sources, which are shown in Figure~\ref{f05}.

\subsection{Spectroscopy}\label{s3-5}
For the \textit{bright} sources, we constructed the background-subtracted spectra and
generated instrumental response files; i.e., redistribution matrix function (RMF) of the
detector and auxiliary response file (ARF) of the telescope. We derived the best-fit
parameters of the spectral models using the $\chi^{2}$ statistics.

For the spectral models, we used an interstellar absorption model (\texttt{tbabs};
\citealt{Wilms00}) convolved with either of the two continuum models: an optically-thin
thermal plasma model (\texttt{apec}; \citealt{smith01}) and a power-law model to
represent the thermal and non-thermal spectra, respectively. Free parameters for the
thermal model are the absorption column (\nh), the plasma temperature (\kt), and the
flux (\fx) in the 0.5--8.0~keV band, while those for the power-law model are the
absorption column (\nh), the photon index ($\Gamma$), and the flux (\fx) in the
0.5--8.0~keV band. For the plasma model, we fixed the abundance to the value in
\citet{gudel07}. We regarded the fitting to be unsuccessful if the reduced $\chi^{2}$
was larger than 1.5 or the best-fit values were unphysical; i.e., $\Gamma>3$ in the
power-law fitting and $\kt>15$~keV in the \texttt{apec} fitting. For sources with
unsuccessful fitting in both models, we also tried a plasma model with two different
temperatures. For sources with successful fitting in both models, we adopted the result
with the smaller reduced $\chi^{2}$ value.

\begin{figure*}[htbp]
 \includegraphics[width=0.3\textwidth]{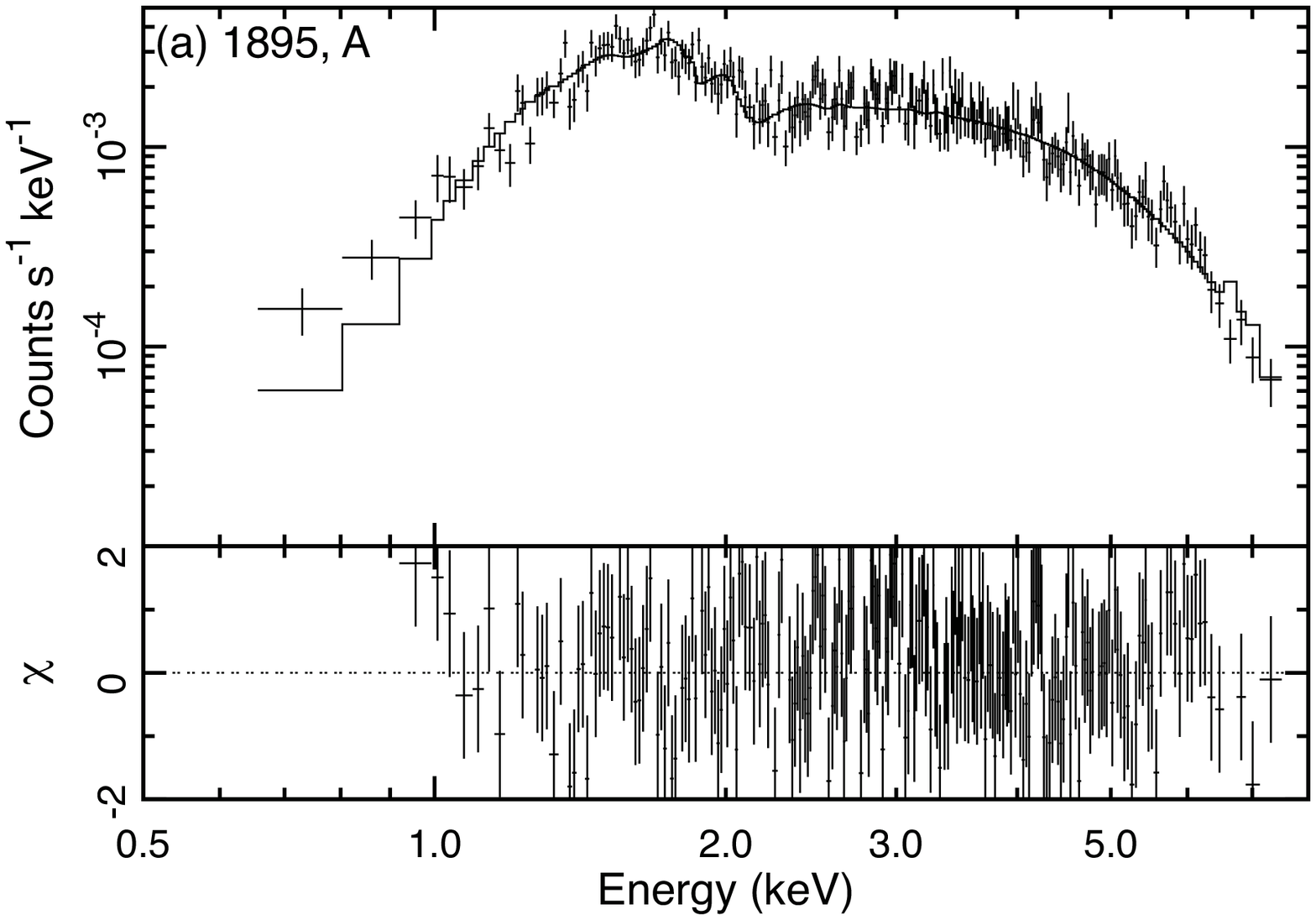}
 \includegraphics[width=0.3\textwidth]{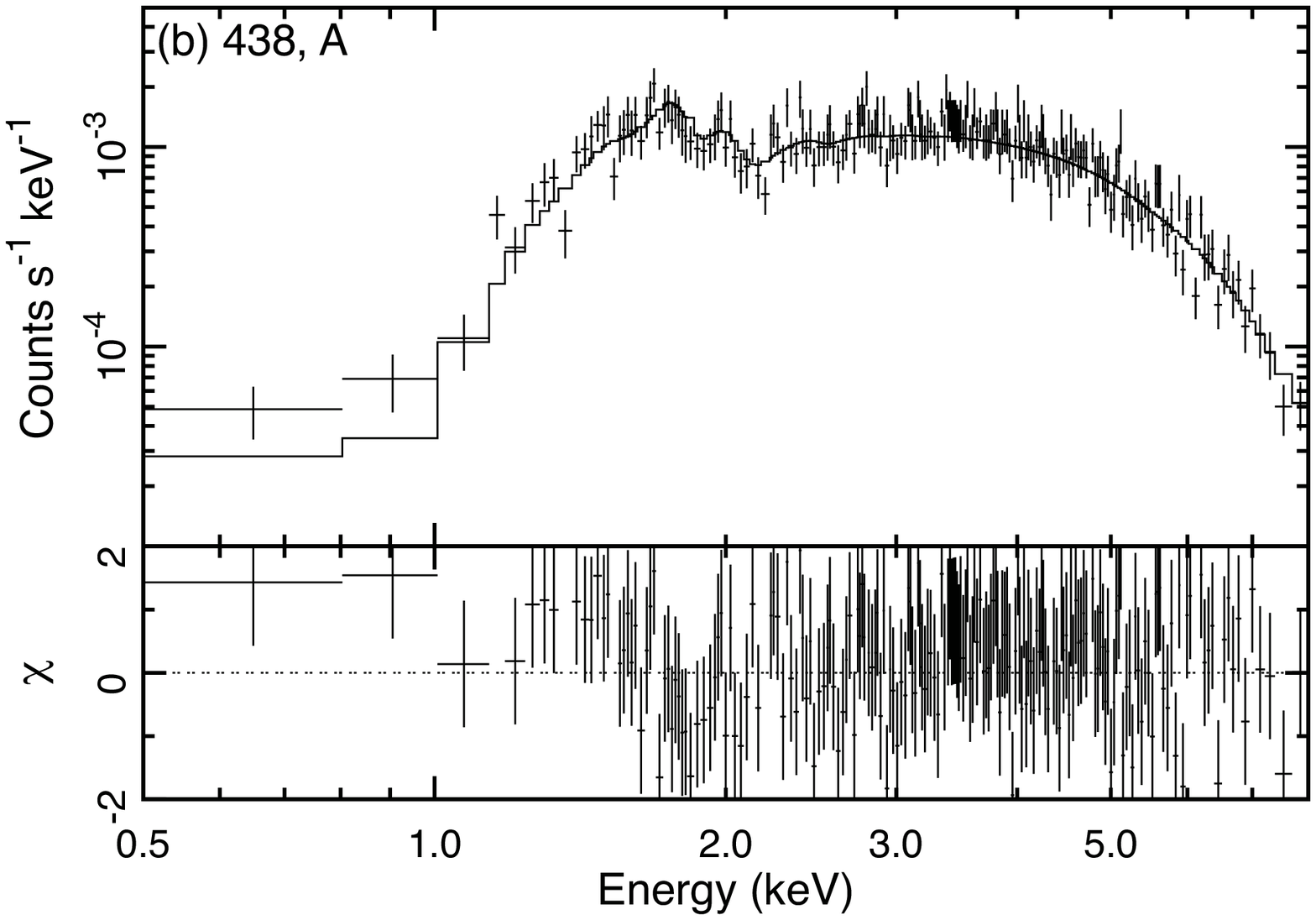}
 \includegraphics[width=0.3\textwidth]{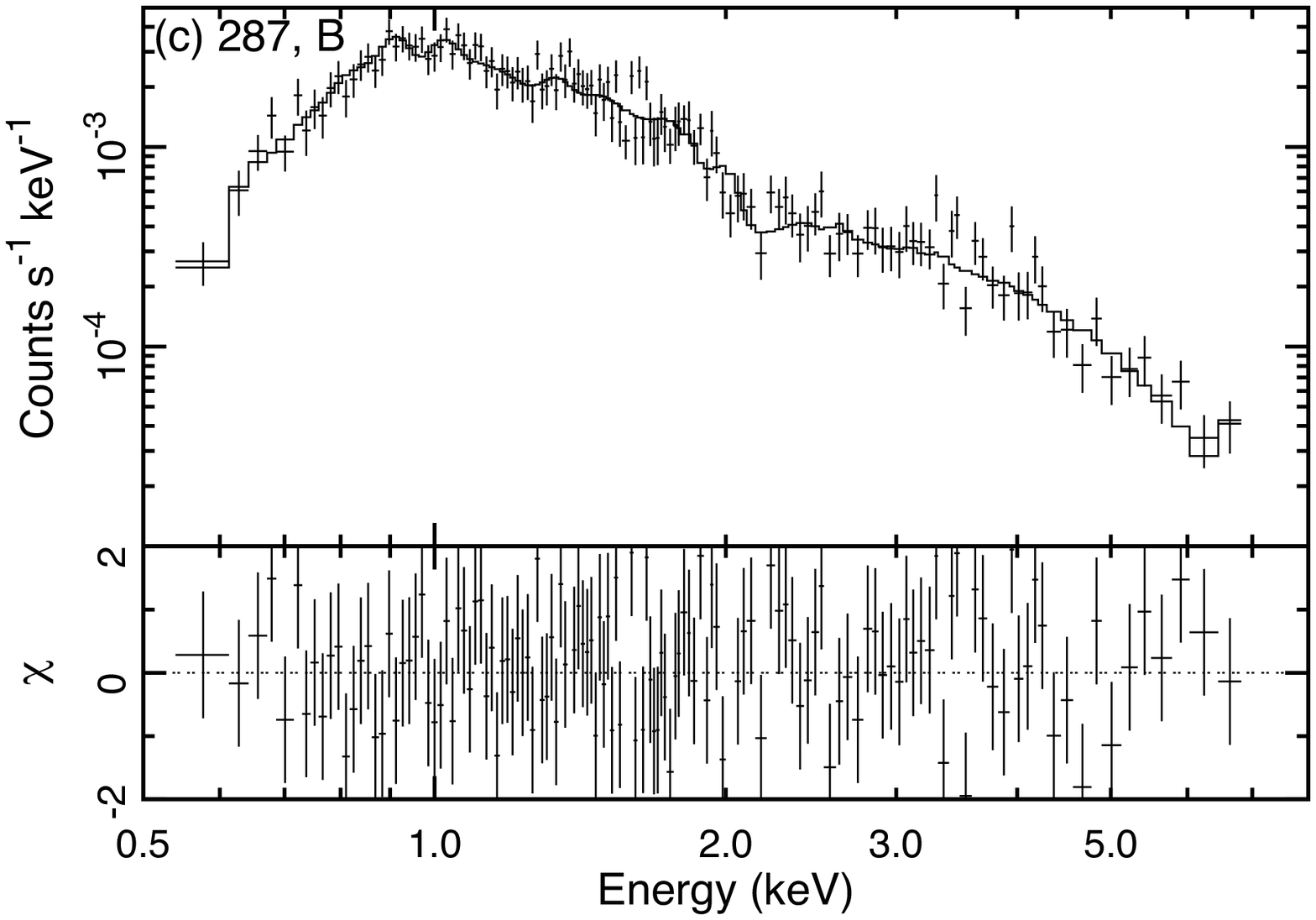}
 \includegraphics[width=0.3\textwidth]{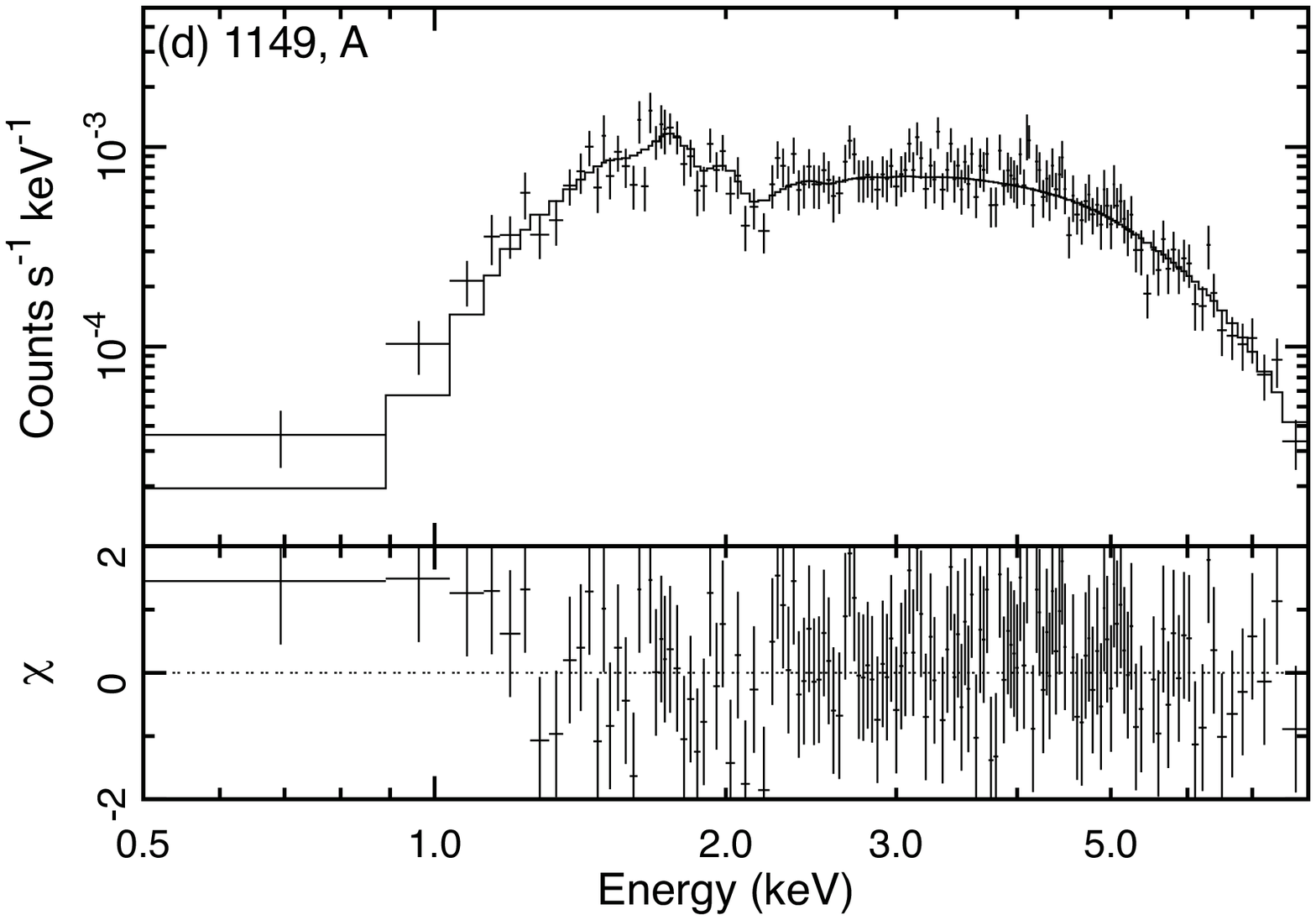}
 \includegraphics[width=0.3\textwidth]{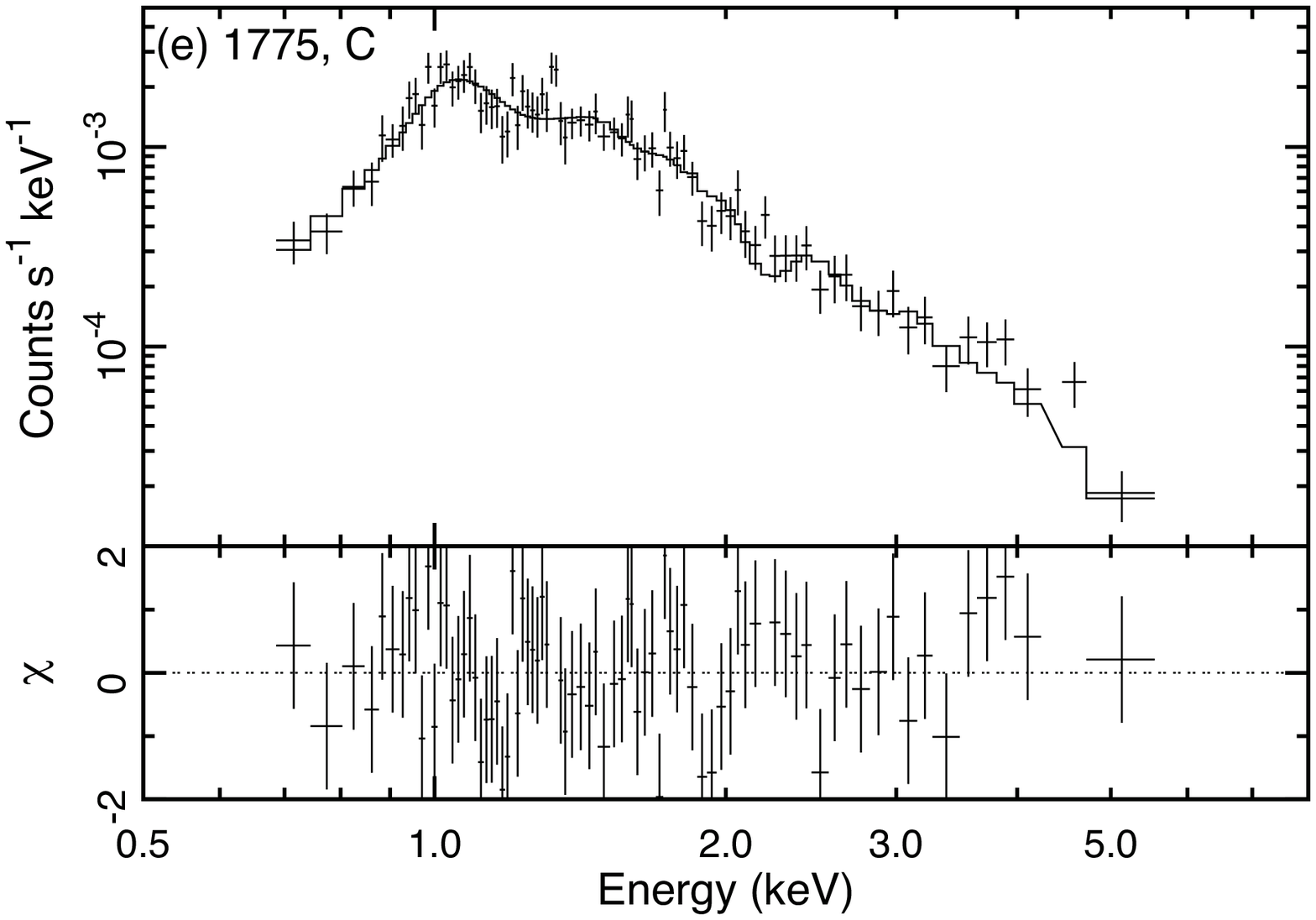}
 \includegraphics[width=0.3\textwidth]{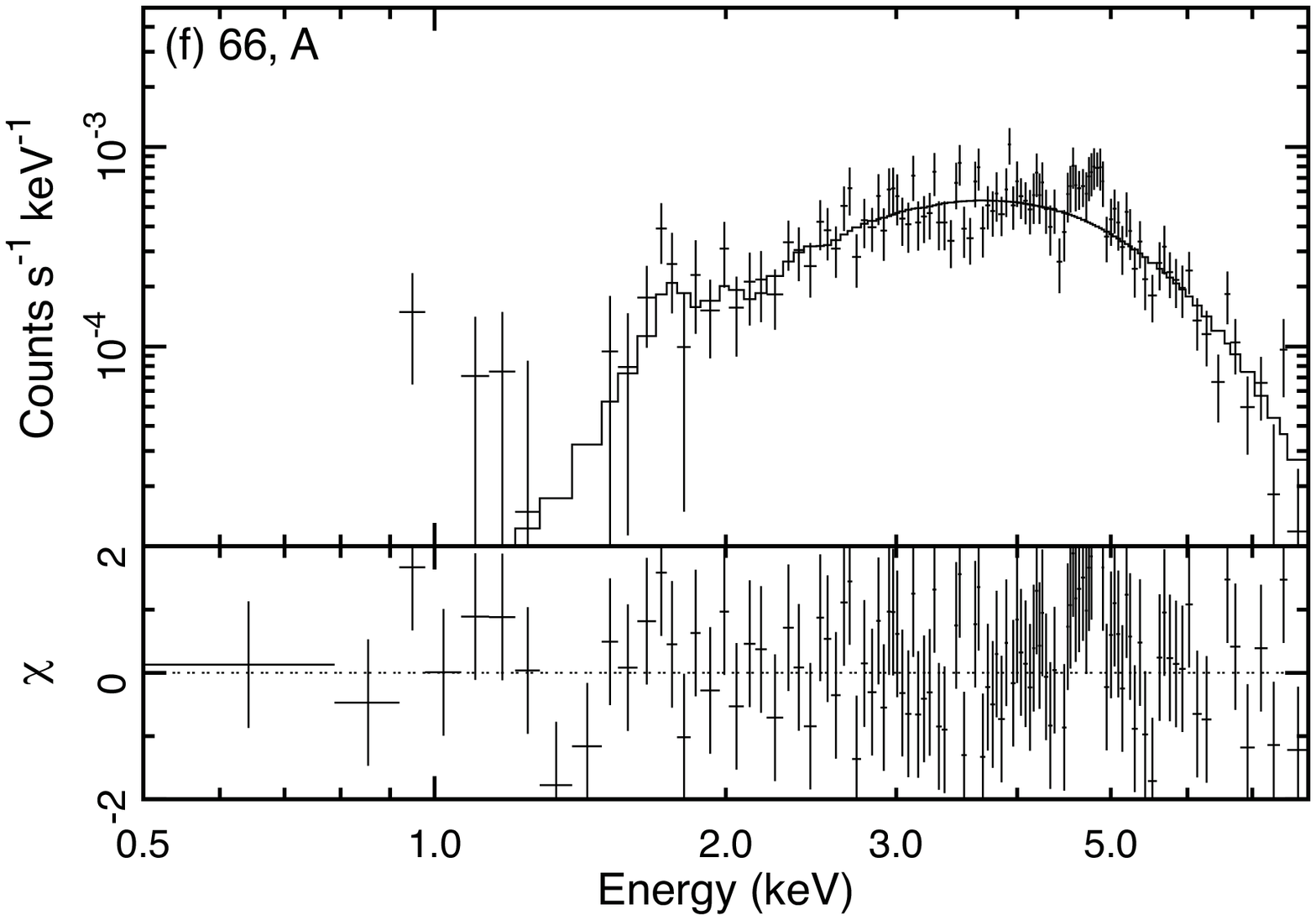}
 \includegraphics[width=0.3\textwidth]{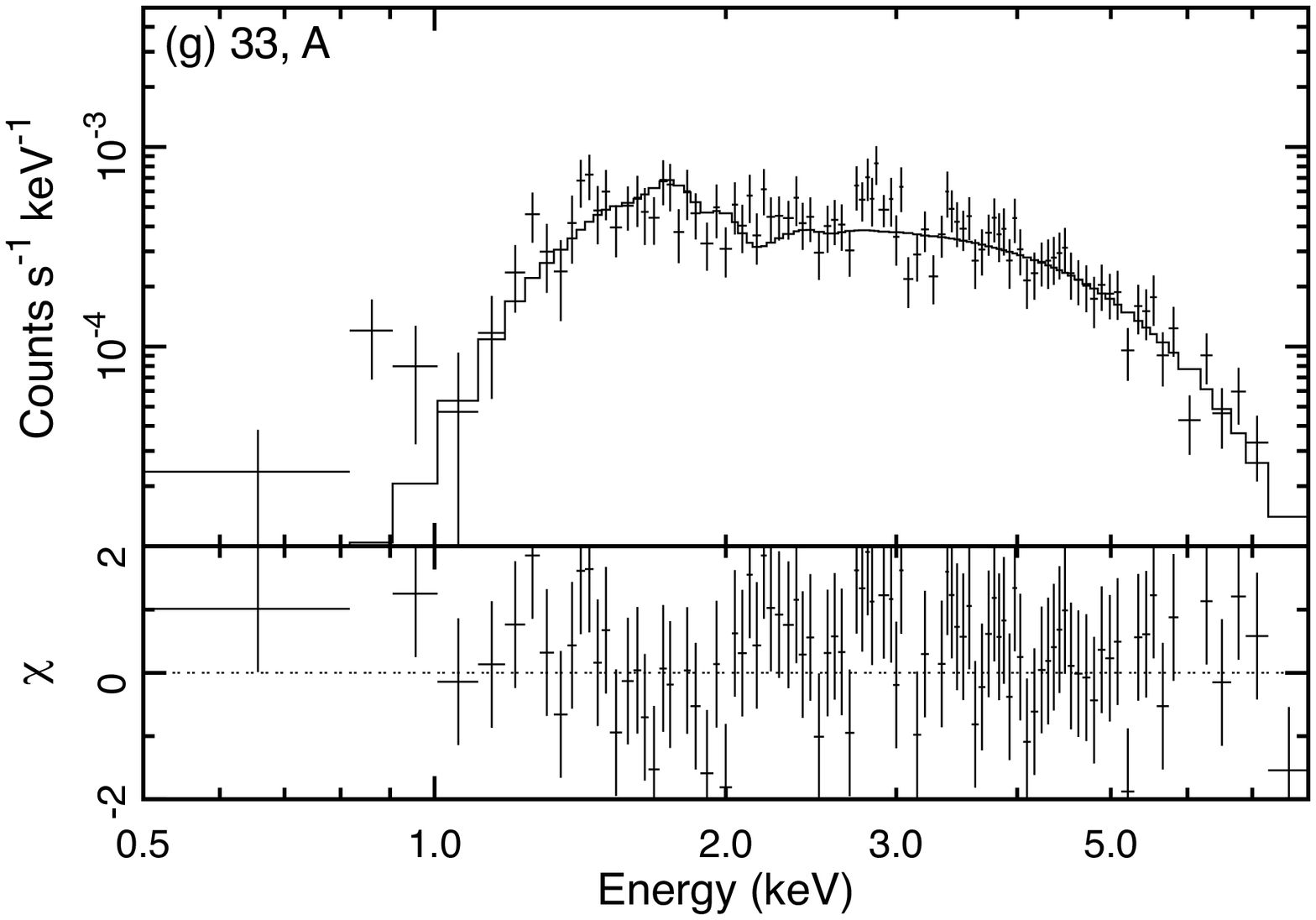}
 \includegraphics[width=0.3\textwidth]{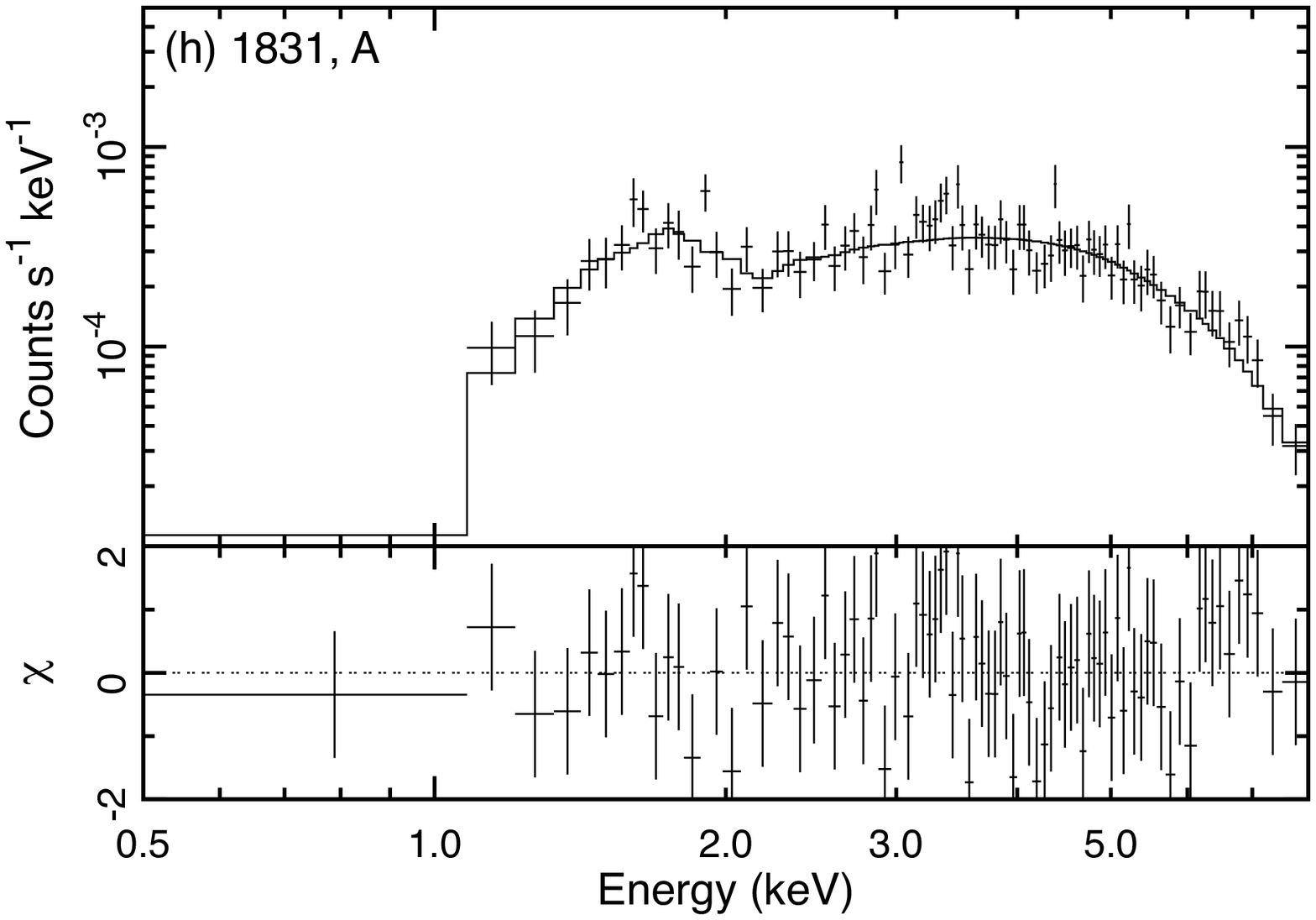}
 \includegraphics[width=0.3\textwidth]{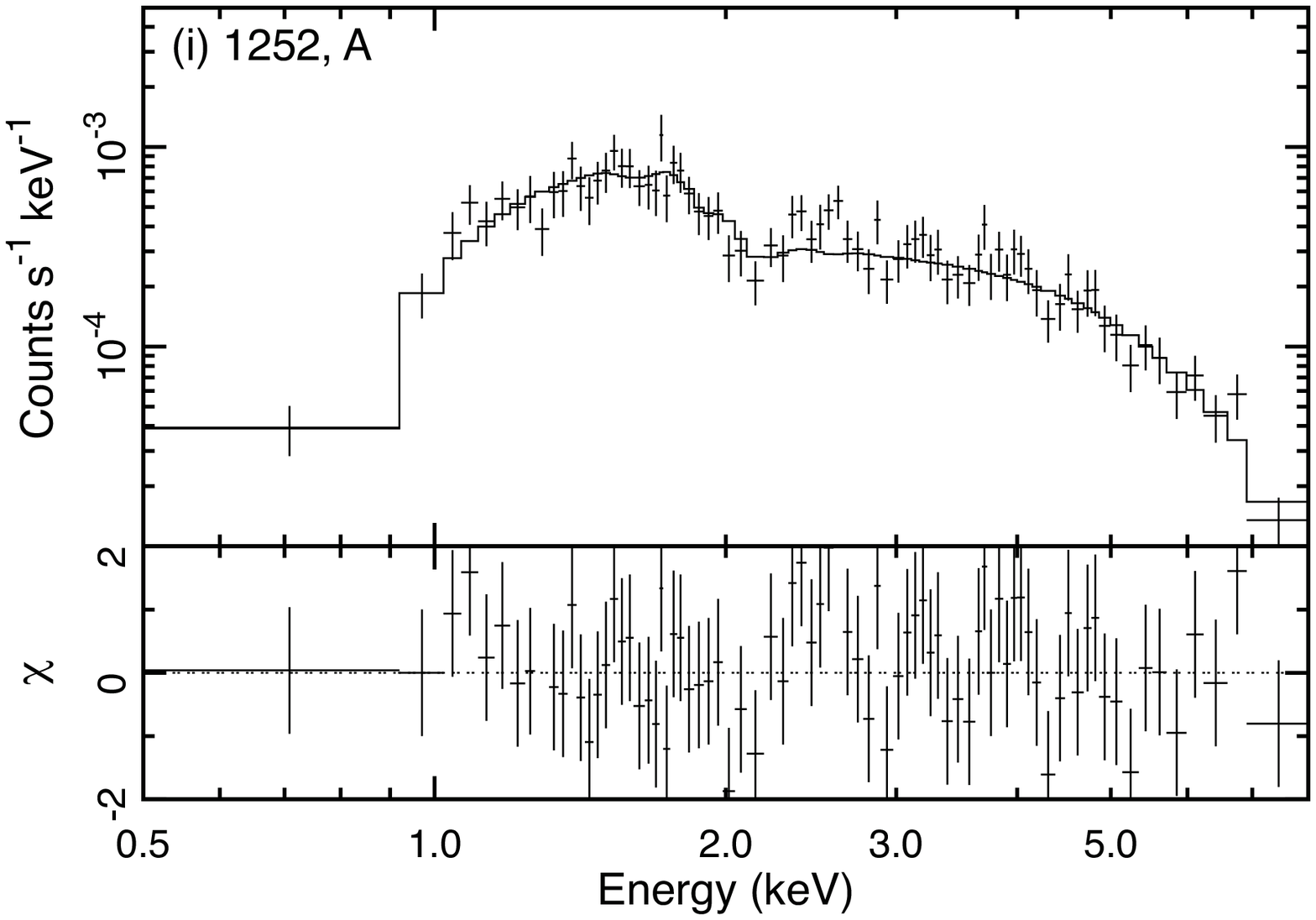}
 \includegraphics[width=0.3\textwidth]{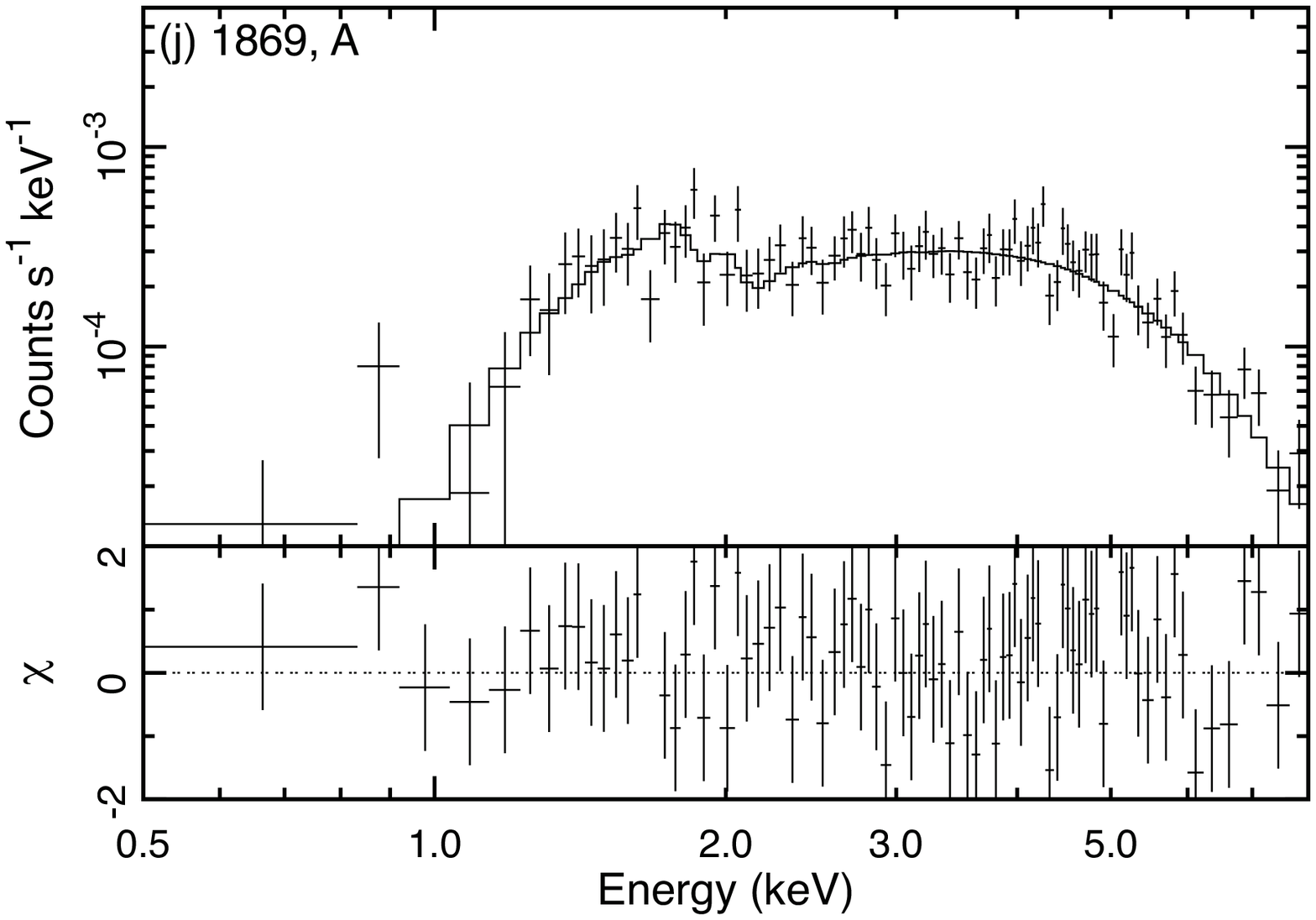}
 \caption{X-ray spectra of the sources with more than 1000 net counts in the order of
 the total counts. The top left labels show the sequence number in Table~\ref{t02} and
 the group (\S~\ref{s4-1}) of each source. Thermal or non-thermal models are employed to
 fit the spectra. Grouped data are shown in the upper panels, which are over-plotted
 with the best-fit model convolved with the instrumental response. The lower panels show
 the residuals of the data to the fit. The best-fit parameters can be found in
 Tables~\ref{t03a} and \ref{t03b}.}
 \label{f06}
\end{figure*}

As a result, 71 sources were successfully fitted with the one-temperature plasma model,
50 sources with the two-temperature plasma model, and 158 sources with the power-law
model. Figure~\ref{f06} shows the spectra and the best-fit models for all sources with
more than 1000 net counts. Tables~\ref{t03a} and \ref{t03b} show the best-fit parameters
for the one-temperature and power-law fitting of these sources, respectively.

\input{t03a}
\input{t03b}

\section{Discussion}\label{s4}
We first attempt to group all the detected X-ray sources in \S~\ref{s4-1} based on the X-ray color-color
diagram. After establishing these groups, we examine the X-ray properties of each group
(\S~\ref{s4-2}). We then discuss likely classes for sources in each group
(\S~\ref{s4-3}) and evaluate their contribution to the GRXE (\S~\ref{s4-4}).

\subsection{Grouping}\label{s4-1}
We constructed an X-ray color-color diagram \citep{hong04} using the quantiles
($E_{25}$, $E_{{50}}$, and $E_{75}$) characterizing the spectral shape of each
source. Here, $E_{x}$ (keV) is the energy below which $x$\% of photons reside in the
energy-sorted event list. $E_{50}$ is equivalent to the median energy. We first
normalized these parameters as
\begin{equation}
 Q_{x}=\frac{E_{\rm{x}}-E_{\rm{min}}}{E_{\rm{max}}-E_{\rm{min}}},
\end{equation}
in which $E_{\rm{min}}$ and $E_{\rm{max}}$ are 0.5 and 8~keV, respectively.  We then
took ratios of two parameters as $q_1 = \log_{10}
Q_{\rm{50}}/\left(1-Q_{\rm{50}}\right)$ and $q_2 = 3Q_{\rm{25}} / Q_{\rm{75}}$ for all
the sources. Here, the $q_1$ value indicates the degree of photon spectrum being biased
toward the higher ($q_1>0$) or lower ($q_1<0$) energy end (hard or soft spectra), and
the $q_2$ value indicates the degree of photon spectrum being less ($q_2>1$) or more
($q_2<1$) concentrated around the peak (broad or narrow spectra).

\begin{figure}[htbp]
 \epsscale{0.6}
 \plotone{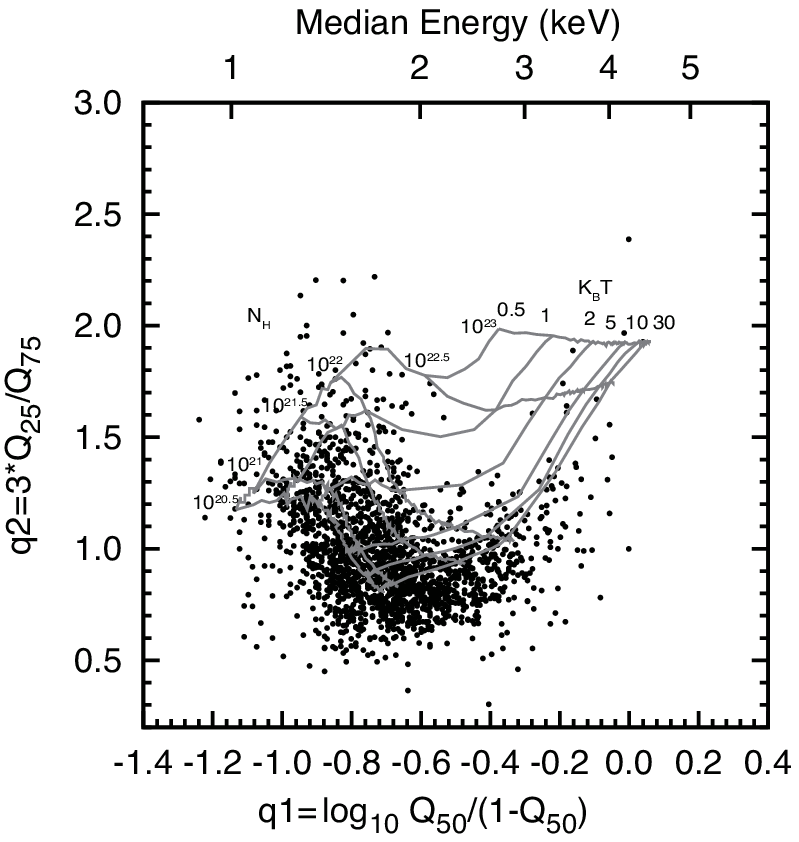}
 \caption{X-ray color-color diagram of all detected X-ray sources. The converted median
 energy is shown in the upper $x$-axis. We also display a grid of thermal plasma spectra
 attenuated by interstellar absorption with various parameter combinations. The labels
 indicate the values of the plasma temperature (\kt) for 0.5, 1, 2, 5, 10, and 30~keV
 and the extinction column (\nh) for 10$^{20.5}$,~10$^{21}$,~10$^{21.5}$, ~10$^{22}$,
 and 10$^{22.5}$~cm$^{-2}$. In the simulation, background events were not included, so
 the grid should be considered only as a reference showing the relative parameter
 changes.  }
 \label{f07}
\end{figure}

Figure~\ref{f07} shows the color-color diagram using $q_1$ and $q_2$ of all the
sources. In order to put the diagram into context, we simulated the quantiles of
optically-thin thermal emission attenuated by interstellar photoelectric absorption. We
used the \texttt{bremss} model for the emission and the \texttt{tbabs} model for the
absorption with varying temperatures and column densities. 

In the plot, sources are distributed from side to side with a main concentration around
($q_1$, $q_2$)~$=$~($-$0.8, 1.0). It appears that the scatter is extended in the upward
and rightward directions from the main concentration. We examine if these branches have
any physical basis by looking at count-stratified color-color diagrams (\S~\ref{s4-1-1})
and the results of \textit{bright} sources (\S~\ref{s4-1-2} and \S~\ref{s4-1-3}) before
dividing all sources into groups by a statistical approach (\S~\ref{s4-1-4}).

\subsubsection{Count-Stratified Diagram}\label{s4-1-1}
\begin{figure*}[htbp]
 \begin{center}
  \includegraphics[width=0.24\textwidth,clip]{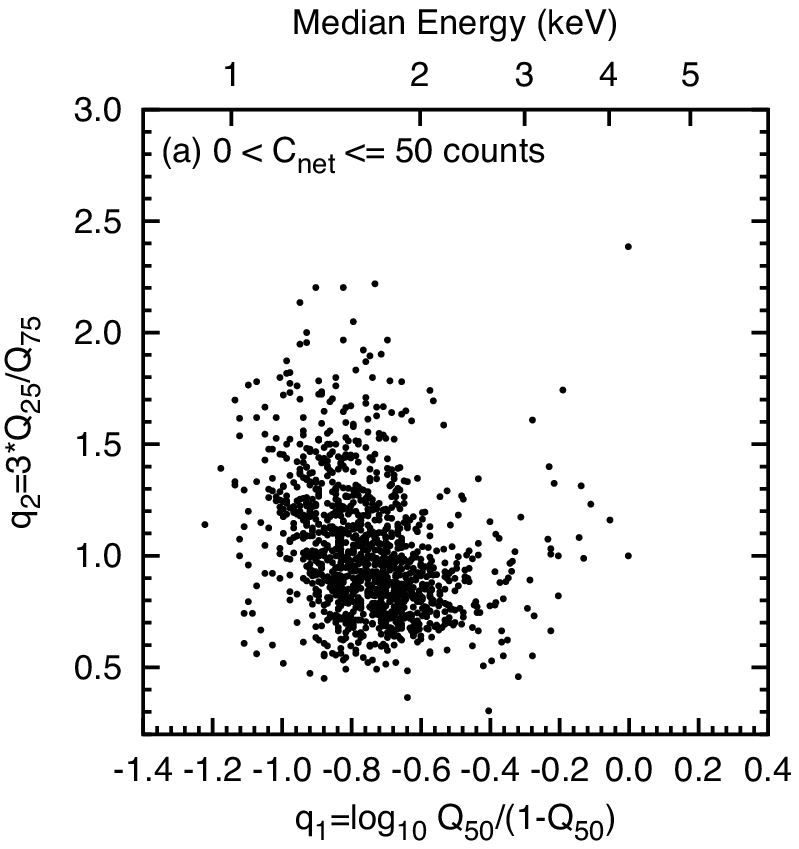}
  \includegraphics[width=0.24\textwidth,clip]{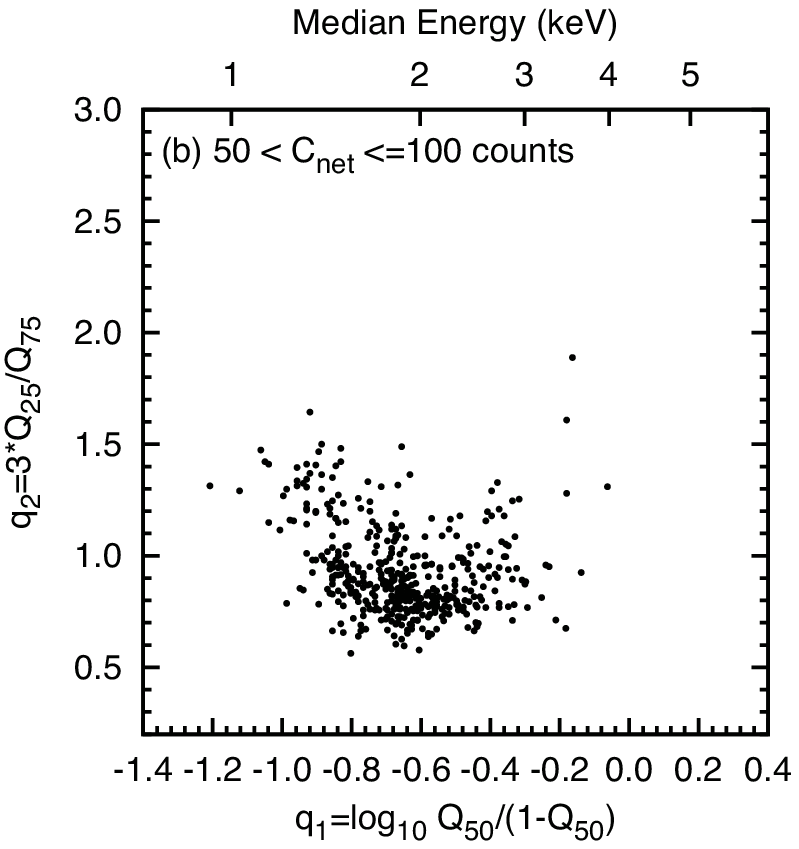}
  \includegraphics[width=0.24\textwidth,clip]{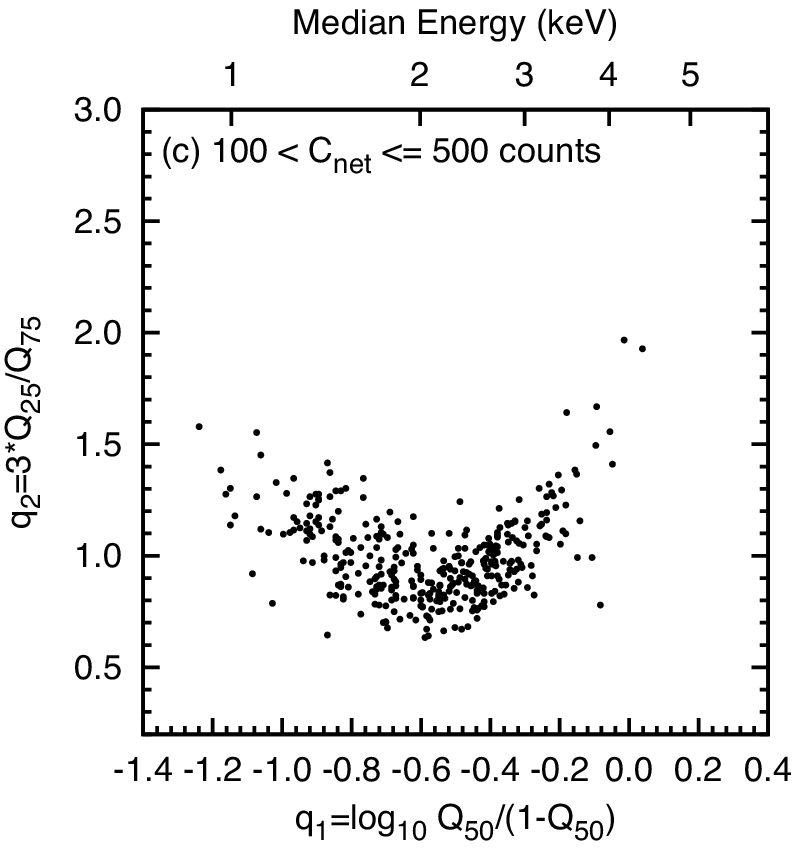}
  \includegraphics[width=0.24\textwidth,clip]{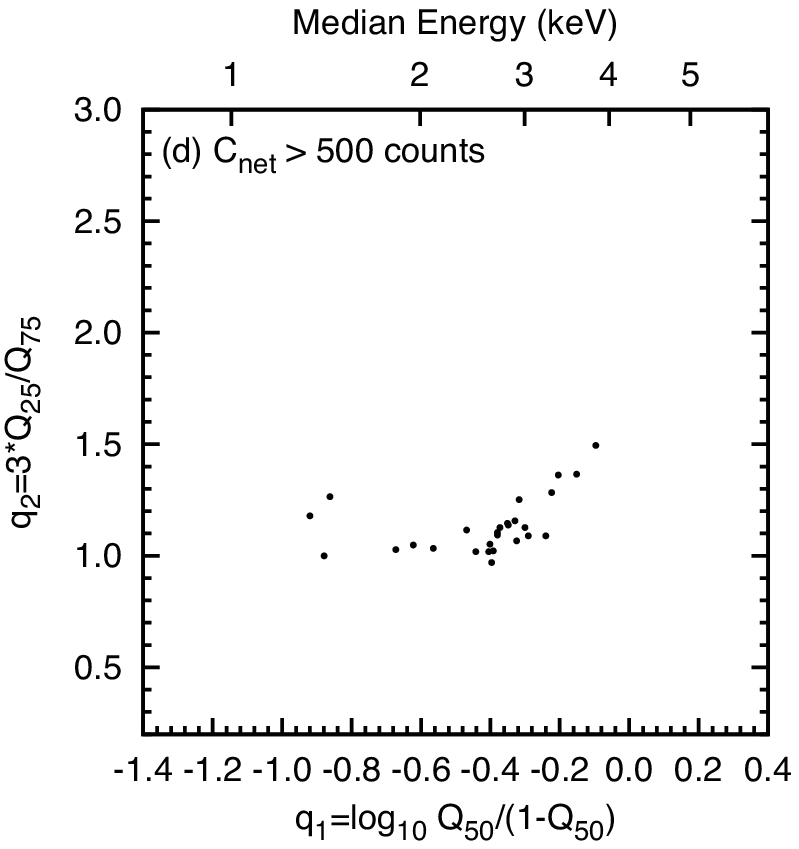}
 \end{center}
 \caption{X-ray color-color diagram for different count ranges. The symbols are the same
 with Figure~\ref{f07}.}
 \label{f08}
\end{figure*}

The color-color diagram does not represent the flux of sources. Therefore, we
constructed the diagram for different count ranges (Fig.\,~\ref{f08}). The scatter plots
exhibit quite a different morphology among the four count ranges. In the lowest count
range (Fig.\,~\ref{f08}a), the sources are scattered in the main concentration and the
upward branch. As the count increases, the sources in the upward branch disappear and
are gradually replaced with sources in the rightward branch (Fig.\,~\ref{f08}b) to form
a \textsf{V}-shaped distribution (Fig.\,~\ref{f08}c). In the highest count range
(Fig.\,~\ref{f08}d), the sources are only seen in the rightward branch.

\subsubsection{Variability of \textit{bright} Sources}\label{s4-1-2}
\begin{figure}[htbp]
 \begin{center}
  \epsscale{0.5}
  \plotone{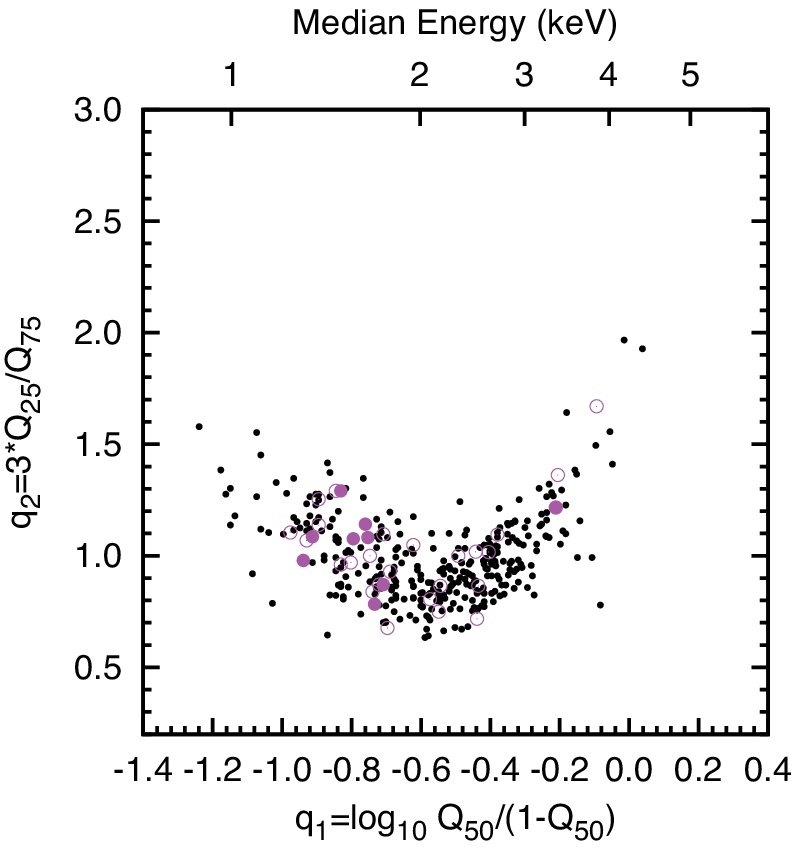}
  \caption{X-ray color-color diagram of the \textit{bright} sources (Cnet $>$ 100 counts). Variable
  sources are shown with magenta circles, among which flare-like variable sources are filled.}
  \label{f09}
 \end{center}
\end{figure}

In the variability analysis of the \textit{bright} sources (\S~\ref{s3-4}), some show
flare-like variability. Figure~\ref{f09} shows where they are located in the color-color
diagram. We omitted the sources with less than 100 counts, as they are too weak to
examine their variability; the faintest source with a statistically significant
variability has a net count of 100.5 (\S~\ref{s3-4}). It is interesting to find that all
but one flare-like variable sources are found along the left arm of the
\textsf{V}-shaped distribution, although the variable sources are equally found in both
arms.

\subsubsection{Spectra of \textit{bright} Sources}\label{s4-1-3}
\begin{figure}[htbp]
 \begin{center}
  \epsscale{0.5}
  \plotone{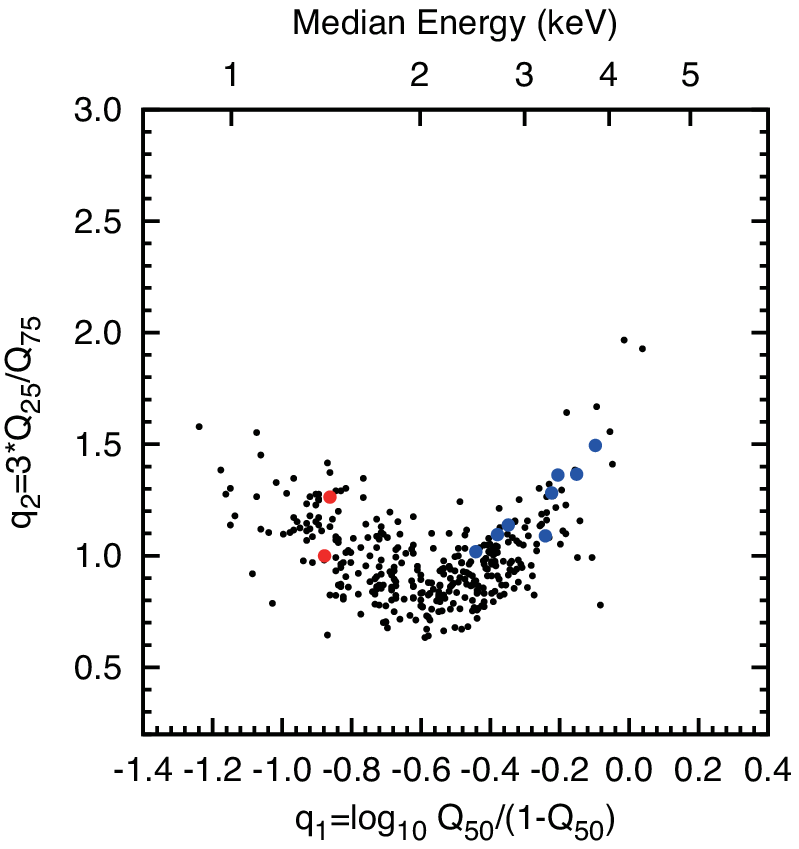}
  \caption{X-ray color-color diagram of the \textit{bright} sources (Cnet $>$ 100 counts). Sources with a
  thermal spectrum (Fig.~\ref{f06}) are shown with red circles, while those with a
  non-thermal spectrum are shown with blue circles.}
  \label{f10}
 \end{center}
\end{figure}

In the spectral analysis of the \textit{bright} sources (\S~\ref{s3-5}), some sources
are described by thermal spectra and others by non-thermal spectra. For sources
with more than 1,000 counts (Fig.\,~\ref{f06}), two sources are thermal and eight are
non-thermal. Figure~\ref{f10} shows where they are located in the color-color
diagram. The bimodality is clearly seen, in which thermal and non-thermal sources are
exclusively found respectively in the left and right arm of the \textsf{V}-shaped
distribution.

\subsubsection{Statistical Treatment}\label{s4-1-4}
From the discussion above, it is reasonable to assume that the sources consist of three
representative groups that can be separated by the position in the diagram: main
concentration, upward branch, and rightward branch. In order to define the boundaries in
an objective manner, we employed the $k$-means clustering algorithm using the
\texttt{mlpy} script\footnote{The details of the script can be found in
http://mlpy.sourceforge.net/.}. Given the number of groups, the algorithm determines a
centroid of each group and the sources that belong to the group iteratively, so that the
sum of the distances from each source to the centroid becomes minimum.

\begin{figure}[htbp]
 \begin{center}
  \epsscale{0.6}
  \plotone{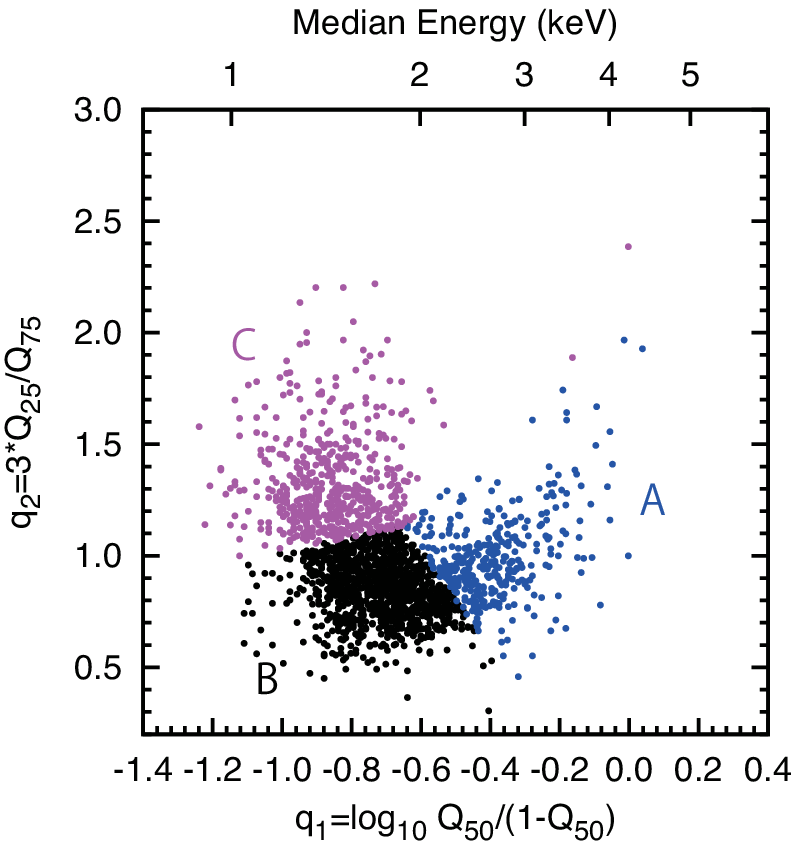}
  \caption{X-ray color-color diagrams of all sources color-coded depending on their
  statistically-defined classes. The other symbols follow Figure~\ref{f07}.}
 \label{f11}
 \end{center}
\end{figure}

Using this method, we divided the 2,002 sources into three groups (group A, B, and C) by
their position in the diagram (Figure~\ref{f11}). The resultant division was mostly
consistent with its appearance; the group A for the rightward branch, B for the main
concentration, and C for the upward branch. Some exceptions are seen; e.g., two sources
at the tip of the rightward branch were divided into the group C, rather than A, which
is conceivable in such an algorithm. The two sources are quite faint comprising only
$<$0.5\% of the total net count of the groups A and C, so they do not affect the
composite spectrum of the groups discussed below.

\subsection{X-ray Properties of Each Group}\label{s4-2}
\subsubsection{Fe K Feature}\label{s4-2-1}
We made composite spectra of the three groups A, B. and C using the CIAO tool
\texttt{combine\_spectra}. The tool makes background-subtracted spectra as well as
count-weighted ARFs and RMFs of multiple sources.

\begin{figure}[htbp]
 \begin{center}
  \epsscale{0.8}
  \plotone{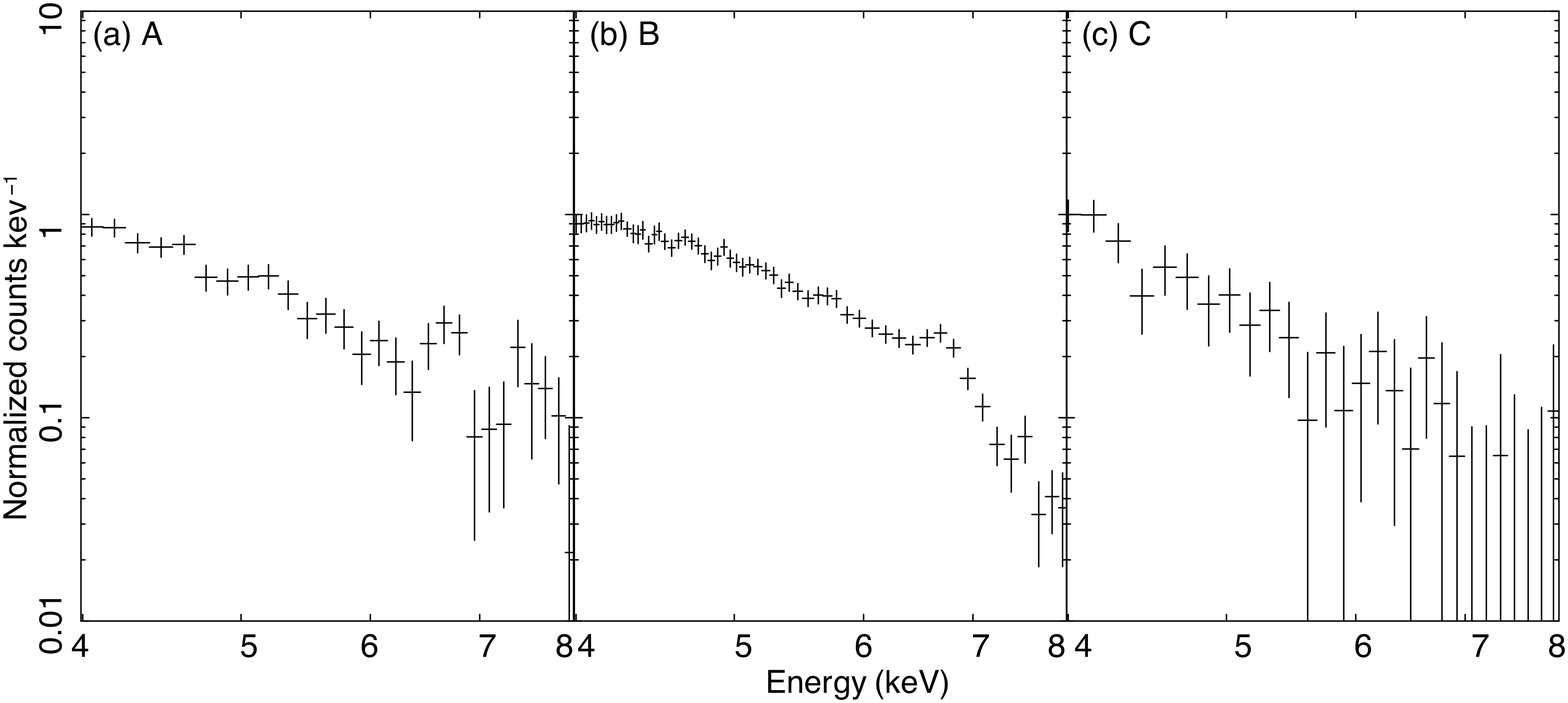}
 \end{center}
 \caption{Fe K feature in the combined spectra of point sources in the group (a) A, (b)
 B, and (c) C. The spectra are normalized at 4~keV.}
 \label{f12}
\end{figure}

Figure~\ref{f12} shows the enlarged view in the 4--8~keV band, in which the Fe K feature
appears differently for each group. We characterized the feature by fitting the 4--8~keV
spectrum using a power-law plus a Gaussian model. The free parameters were the power-law
index and normalization, and the Gaussian line center, width, and normalization. For the
group C, the Gaussian component was not required, so we only constrained the upper limit
to the emission. The result is shown in Table~\ref{t05}. For the best-fit value of the
Gaussian line center, we found no significant differences among different groups.

\input{t05}

The characterization of the Fe K feature was conducted also for the composite spectrum
of all point sources and the total emission extracted from the entire field. The
fractional contribution of the continuum (4--8~keV) and the Fe K band emission was
derived separately for each group. The fraction of point sources among the entire
emission is smaller than that reported ($\sim$80\%) in \citet{revnivtsev09}, which is
presumably due to our lower sensitivity on average by using the entire field rather than
the central part of the observations.

\medskip

The presence of the Fe K feature in the combined spectrum of the group A sources
(Fig.~\ref{f12}a) seemingly contradicts the result of the spectral fitting of individual
\textit{bright} sources (\S~\ref{s3-5}), because most of the \textit{bright} sources in
the group A hardly show signatures of the Fe K feature. Among the brightest ten sources
shown in Figure~\ref{f06}, eight are group A sources (source numbers 1895, 438, 1149,
66, 33, 1831, 1252, and 1869), all of which are fitted with a power-law model
(Table~\ref{t03b}). This strongly suggests that the presence of the Fe K feature depends
on the source flux within the group A. No such apparent contradiction was found in the
other groups.

In order to investigate the flux dependence for the Fe K contribution, we constructed
spectra stacking flux-sorted sources cumulatively from the brightest source toward the
fainter end by increasing the source number by 1 (1--20), by 10 (20--40), and by 20
(40--) at each step. In other words, the composite spectrum was made for the brightest
1, 2, ..., 20, 30, 40, 60, 80, ... sources. We then applied a power-law plus a Gaussian
model for the spectrum generated at each step to derive the EW of the Fe K
feature. Figure~\ref{f12} (a) shows the EW value against the decreasing flux. At the
brightest end ($>$2$\times$10$^{-14}$ $\ergcms$), the EW is consistent with being zero,
which gradually increases as the threshold flux decreases and eventually levels
off. This suggests that different classes of sources are dominant in the brighter and
fainter end of the flux and their ratio gradually changes along the flux.

\begin{figure}[htbp]
 \epsscale{0.6}
 \plotone{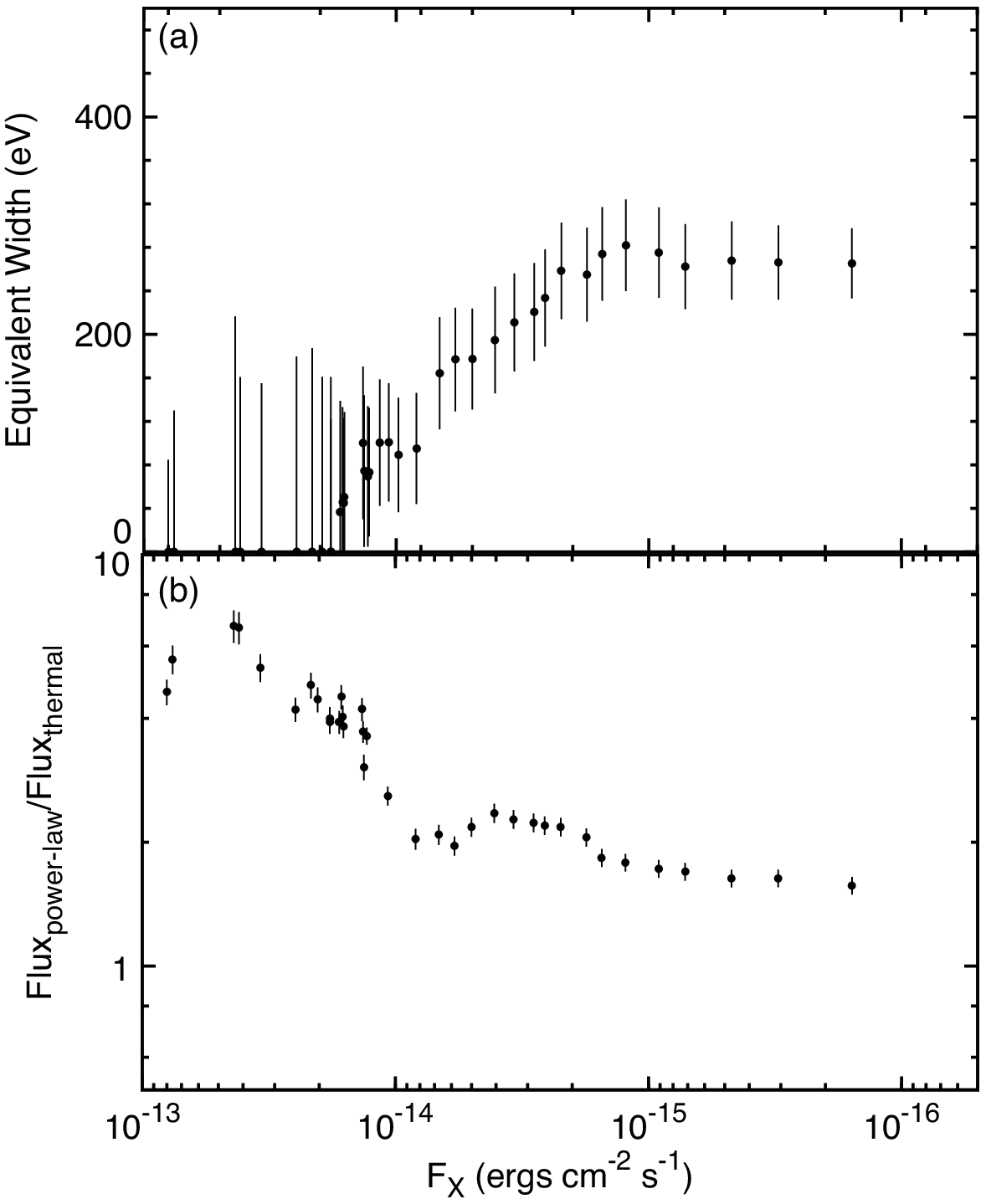}
 \caption{(a) EW of the Fe K feature and (b) the 2--8~keV flux ratio of the power-law
 ($F_{\mathrm{power-law}}$) versus thermal ($F_{\mathrm{thermal}}$) components against
 the decreasing flux in 2--8~keV above which the cumulative combined spectra were
 constructed. The 1$\sigma$ statistical uncertainty is shown for each data point.}
\label{f13}
\end{figure}

\subsubsection{Global Spectral Fitting}\label{s4-2-2}
We performed global spectral fitting using the entire energy band (0.5--8.0~keV) for the
composite spectrum for each group (Fig.~\ref{f11}) in a similar manner for the
individual \textit{bright} sources (\S~\ref{s3-4}). For the thin-thermal plasma model,
we fixed the abundance to that adopted in \citet{gudel99}.

\begin{figure*}[htbp]
 \begin{center}
 \includegraphics[width=0.32\textwidth]{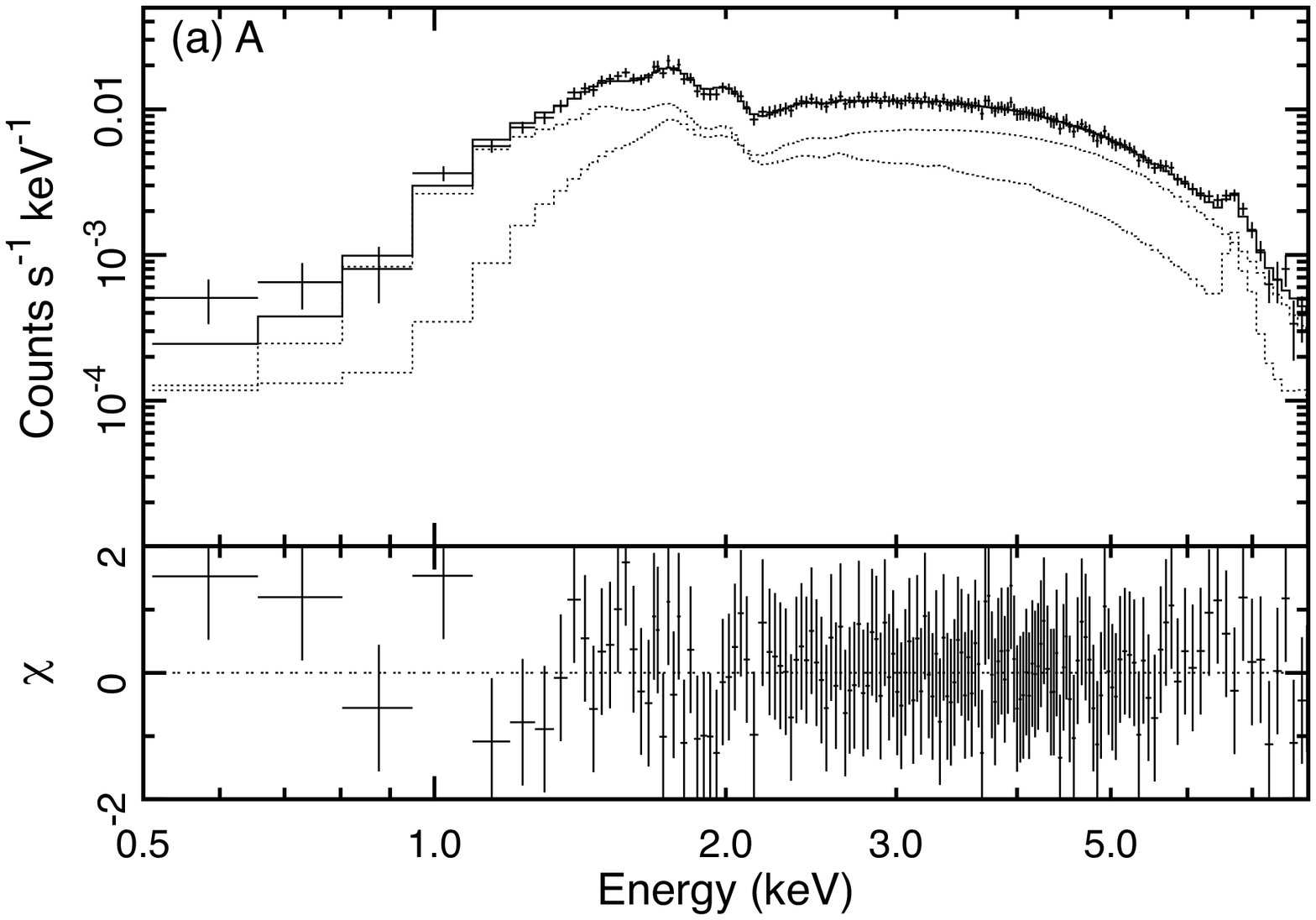}
 \includegraphics[width=0.32\textwidth]{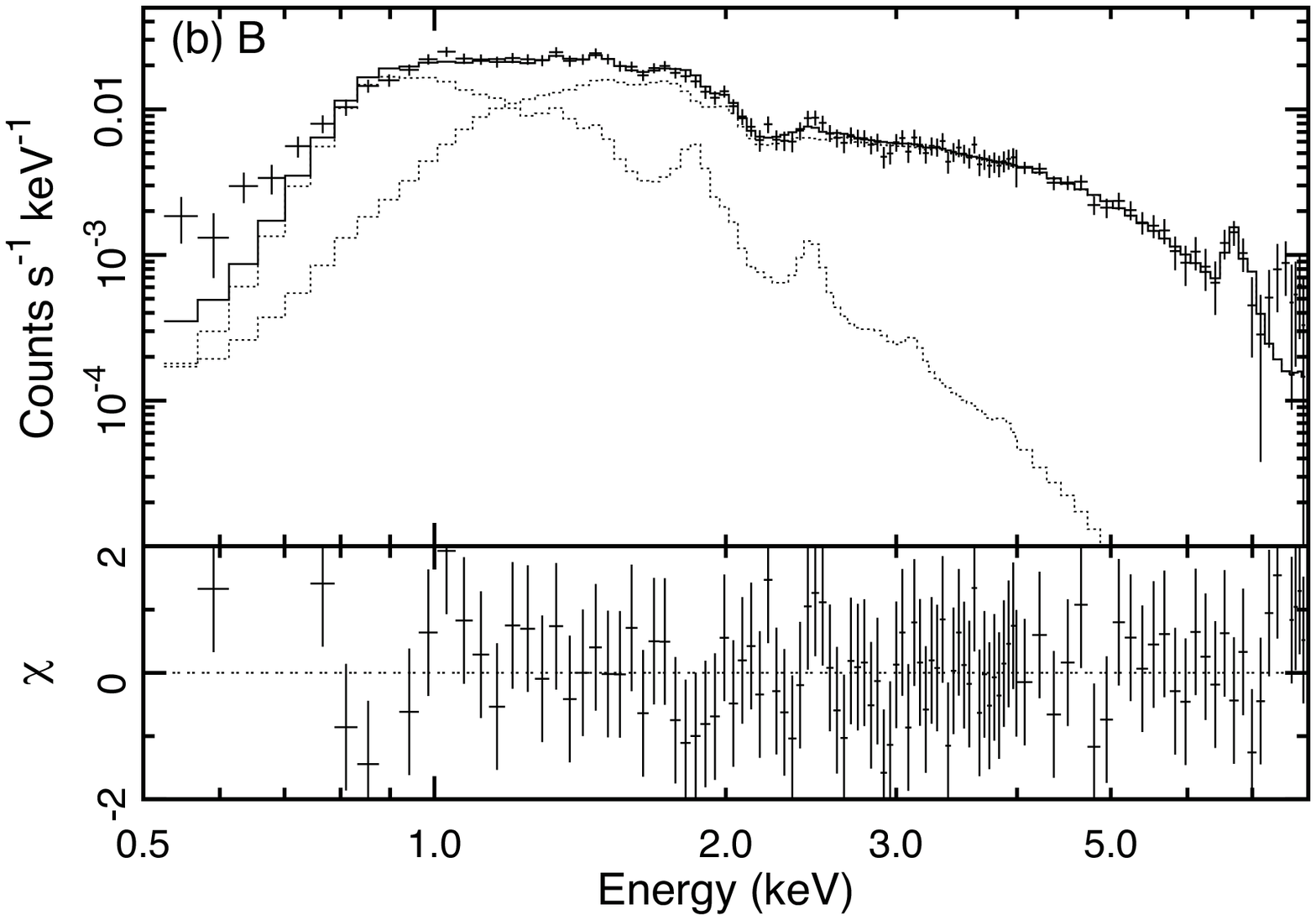}
 \includegraphics[width=0.32\textwidth]{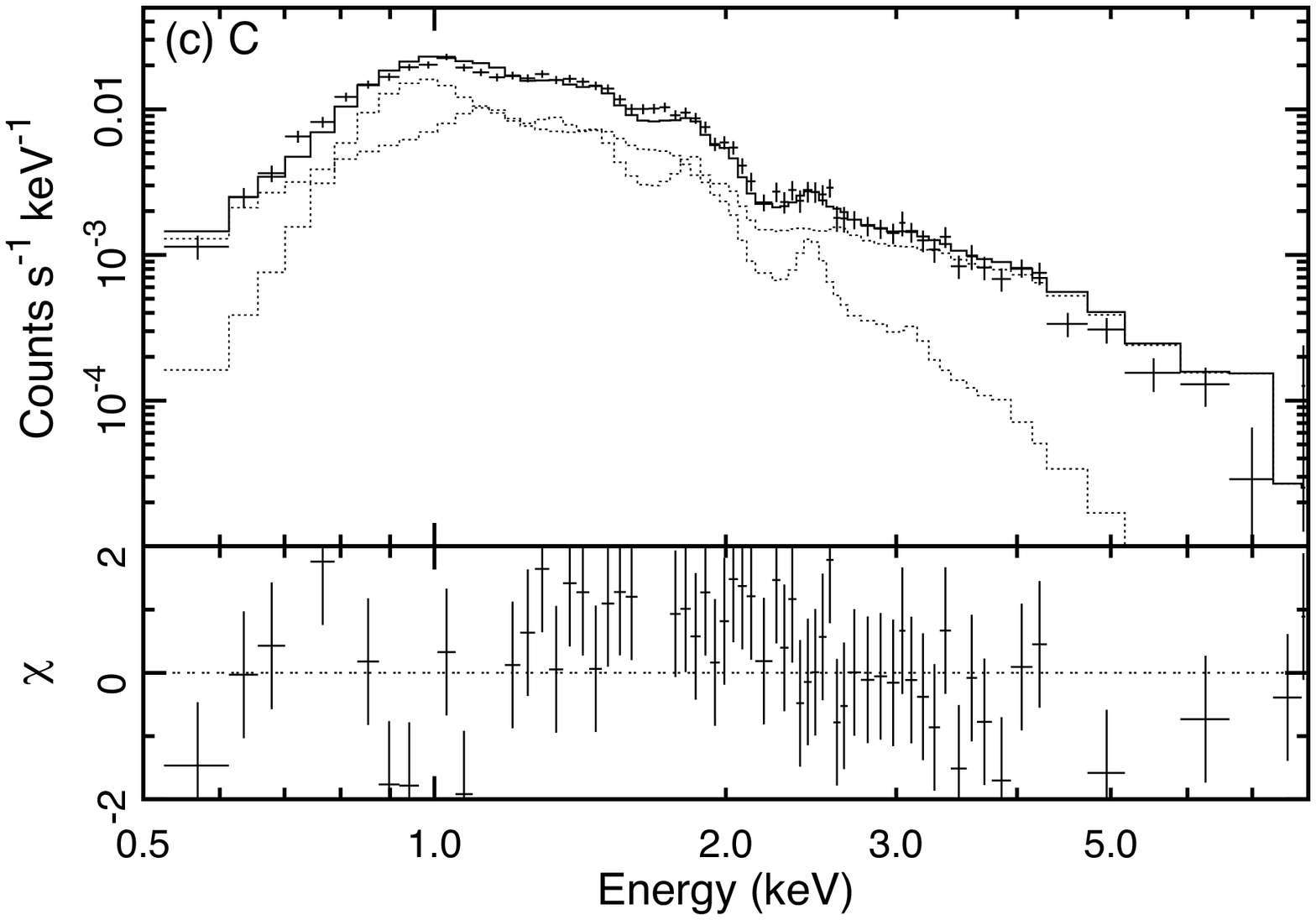}
 \end{center}
 \caption{Composite spectra and the best-fit global model of the group (a) A, (b) B, and
 (c) C. The symbols follow Figure~\ref{f06}.}
\label{f14}
\end{figure*}

\paragraph{Group A}
For the group A spectrum, neither a power-law nor a thin-thermal plasma model did not
reproduce the spectrum well respectively because of the excess emission at 6.7~keV or a
flatness of the continuum. In fact, these two are a signature respectively of a
thin-thermal and a power-law model, so we fitted the spectrum with a combination of the
two, which was successful. The best-fit model are given in Figure~\ref{f14}, whereas the
best-fit parameters are in Table~\ref{t06}.

The EW of the Fe K feature becomes larger as the flux decreases (\S~\ref{s4-2-1}), which
suggests that the thermal component becomes stronger against the non-thermal component as
the flux decreases. In order to investigate this trend, we fitted the cumulatively
stacked spectra (\S~\ref{s4-2-1}) with a power-law plus an optically-thin thermal plasma
model. The free parameters are the normalization of the two components, and the other
parameters were fixed to the values obtained in the fitting of all group A source
(Table~\ref{t06}). As is expected, the flux ratio of the two components
$F_{\mathrm{power-law}}$/$F_{\mathrm{thermal}}$ starts to decrease at
$\sim$10$^{-14}$~$\ergcms$ (Fig.\,\ref{f13}b), which is consistent with the EW trend
(Fig.\,\ref{f13}a).

\paragraph{Group B \& C}
For the group B spectrum, several emission lines are seen, including the 6.7~keV
emission from \ion{Fe}{25} and 2.5~keV from \ion{S}{15}. This set of emission lines
indicates a multiple-temperature plasma, and indeed the spectrum was reproduced well with
two thin-thermal plasma components, but not with one component. The free parameters are
the plasma temperature, metallicity, and absorption column. The metallicity was the same
among all the metals and was constrained to be the same between the two plasma
components.

For the group C spectrum, several emission lines are also seen. Unlike the group B
sources, the \ion{Fe}{25} emission at 6.7~keV is absent.  We fitted the
spectrum using the same model with group B and obtained the best-fit parameters in
Table~\ref{t06}.

\input{t06}

\subsubsection{The $\log N$--$\log S$ Curve}\label{s4-2-3}
\begin{figure}[htbp]
 \begin{center}
  \epsscale{0.8}
  \plotone{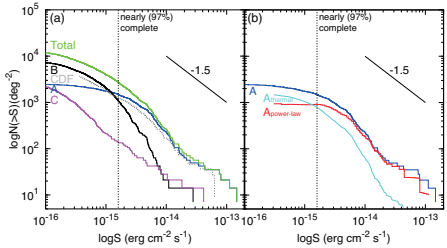}
  \caption{Cumulative source surface number density $N$ (deg$^{-2}$) against the flux
  $S$~$\ergcms$ in the hard band (2--8~keV) separately for (a) all point sources (green)
  and those belonging to the group A (blue), B (black), and C (magenta), and (b) those
  decomposed into power-law (A$_{\mathrm{power-law}}$; red) and thermal
  (A$_{\mathrm{thermal}}$; cyan) sources. The curve for extra-galactic sources with an
  absorption of $\sim$10$^{22}$ cm$^{-2}$ is shown with the gray dotted curve using the
  CDF data \citep{Rosati02} in (a). The slope of --1.5 for spatially uniform
  distribution is shown at the top right. All curves are not corrected for the detection
  completeness. The nearly (97\%) completeness limit is shown with the vertical dashed
  line.}
  \label{f15}
 \end{center}
\end{figure}

We made the $\log N$--$\log S$ curve separately for all point sources and those for each
group (solid histograms in Fig.~\ref{f15}). The curves were constructed in the hard
band. For the group A curve, we divided the cumulative number into two, representing the
power-law and plasma sources by using the smoothed flux ratio curve as a function of
flux (Fig.~\ref{f13}b). We did not correct the $\log N$--$\log S$ curve because we do
not discuss the absolute value of the vertical axis of the $\log N$--$\log S$ curve.

It is expected that a certain fraction of point sources are extra-galactic sources,
which we represent with the $\log N$--$\log S$ curve constructed in a \chandra~deep
field (CDF; \citealt{Rosati02}). Since the extra-galactic sources in our study field are
more absorbed than those in the CDF at a high Galactic latitude, we took into account
the flux reduction due to the Galactic absorption with a column density of
10$^{22}$~cm$^{-2}$ assuming a typical spectral shape of sources in the CDF
\citep{tozzi06}. The additional Galactic column density is a typical best-fit value
among the sources with spectral fitting (Table~\ref{t03a}).

\subsection{Likely Populations}\label{s4-3}
We now discuss likely classes of the X-ray point sources in each group based on the
results presented above. The transition from one class to another is continuous along
the color and flux, so the groups are naturally a mixture of sources of different
classes. We thus discuss the likely classes of sources that constitute the majority of
each group.

First, we consider that the group A sources are mostly a mixture of AGNs and WD
binaries, each responsible for the power-law and thermal plasma component in the
composite spectrum (Fig.\,\ref{f14}a). For the power-law component of the group A, the
$\log N$--$\log S$ curve (Fig.~\ref{f15}) matches quite well to the completeness limit
(Fig.~\ref{f02}) with that obtained in the CDF \citep{Rosati02}. In the spectral
fitting, the power-law component has a spectral index of 1.29$^{+0.18}_{-0.40}$
(Table~\ref{t06}), which is similar to the typical spectral form of AGNs with a
power-law index of about 1.63$\pm$0.13 in the X-ray band regardless of their types and
luminosities (e.g., \citealt{Rosati02,charles97}).

For the thermal component of the composite spectrum (Fig.~\ref{f14}a) of the group A, a
strong Fe K feature and a 7~keV temperature plasma (Table~\ref{t06}) are seen. Both of
these features are observational characteristics of magnetic CVs \citep{ezuka99}. Other
classes of white dwarf (WD) binaries, such as dwarf novae, pre-CVs, and symbiotic stars,
may also be considered for the likely classes. Pre-CVs are poorly recognized class of
sources, which are detached binaries of a WD and a late-type star, unlike conventional
CVs that are semi-detached systems. Some of them show strong Fe K emission in the hard
X-rays \citep{matranga12}. Likewise, some symbiotic stars are also known to show strong
Fe K emission feature \citep{eze12}. In fact, \citet{morihana12} showed near-infrared
spectra of selected X-ray sources presented in this paper, in which some thermal A
sources do not exhibit the Br$\gamma$ emission that is typical for the conventional CVs
\citep{dhillon97}. We thus consider that pre-CVs and symbiotic stars, most of which do
not show the Br$\gamma$ emission \citep{Howell10, Schmidt03}, also account for at least
some fraction of the thermal source population in the group A.

The group B and C sources are Galactic sources with a soft thermal spectrum, and we
consider that most of them are likely to be X-ray active stars. The composite spectra of
the two groups were fitted with two plasma components (Fig.~\ref{f14} and
Table~\ref{t06}). The lower temperature component and its absorption column are quite
similar between the two, indicating that the difference between the B and C are not
their typical distances. On the contrary, the higher temperature component is
significantly higher for the group B; it is is high enough to produce a strong Fe K
feature at 6.7~keV, which is absent in the C spectrum. We speculate that the difference
between B and C is that most C sources represent active binary stars in the quiescence,
while most B sources represent those during flares. Although most flare-like variable
sources are indeed group B sources (Fig.\,~\ref{f09}), this can be a statistical bias
because the group C sources are systematically fainter than group B sources
(Fig.\,~\ref{f08}).

\subsection{Comparison with Previous Studies and Contribution to GRXE}\label{s4-4}
We have constructed a $\log N$--$\log S$ curve in the 2--8~keV band down to the flux
$\sim$10$^{-16}$ $\ergcms$ using the CBF data (\S~\ref{s4-2}). Several major groups of
sources account for different fractions of the GXRE continuum and Fe K$\alpha$
emission in different flux ranges.

For the continuum emission, low-mass X-ray binaries dominate in the flux range above
$\approx$10$^{-12}$~$\ergcms$ in the $\log N$--$\log S$ curve, as was found in previous
studies (e.g., \citealt{hands04}). These bright Galactic sources saturates, and
background AGNs emerge as the most dominant class of sources in the next flux range of
$\approx$10$^{-12}$--10$^{-14}$~$\ergcms$. The dominance of AGNs over Galactic
population in this flux range was also discussed by
\citet{ebisawa01,ebisawa05,hands04,revnivtsev09,motch10}. Below
$\approx$10$^{-14}$~$\ergcms$, Galactic sources start to dominate the $\log N$--$\log S$
curve again. We decomposed the $\log N$--$\log S$ curve of the group A sources into
non-thermal (AGNs in our speculation) and plasma sources (WD binaries) by utilizing the
increasing EW of the Fe K$\alpha$ emission in the composite spectra along the decreasing
flux. The group A thermal sources (WD binaries) and the group B sources (X-ray active
stars) are the major contributers in this flux range. These two Galactic classes are
also proposed to be dominant in this flux range by \citet{revnivtsev09,hong12}. The
former study claimed that the X-ray active stars account for more than a half of the
integrated flux, while the latter study claimed the opposite. In our study, the
integrated number of sources of the group A thermal sources (WD binaries) are taken over
by that of the group B sources (X-ray active stars) at
$\sim$3$\times$10$^{-15}$~$\ergcms$ (Fig.~\ref{f15}).

For the Fe K$\alpha$ emission, we found that the group A thermal sources (WD binaries)
and the group B sources (X-ray active stars) are the two major contributers
(Table~\ref{t05}), despite their much smaller contributions to the continuum emission in
comparison the group A non-thermal sources (AGNs). If we assume that all the Fe K$\alpha$
emission of the group A sources (Fig.~\ref{f14}a) is from its
thermal component, the fractional contribution to the Fe K$\alpha$ emission by the
group A thermal (WD binaries) and group B sources (X-ray active stars) is about 2:1
(Table~\ref{t05}).

\section{Summary}\label{s5}
We presented the result of coherent photometric and spectroscopic analysis of 2,002 X-ray
point sources using the entire data set of the deep \textit{Chandra} bulge field
observations, based on which we discussed X-ray source population in the hard X-ray
band in the flux range between $\approx$10$^{-13}$~$\ergcms$ and
$\approx$10$^{-16}$~$\ergcms$ contributing to the GRXE. The main results are
as follows:

\begin{enumerate}
\item Based on the X-ray photometric information, we phenomenologically classified all
      the sources into three groups A, B, and C. The group A sources are the mixture of
      two different classes of a thermal and non-thermal spectrum. From their X-ray
      properties, we speculated that the most dominant class of these groups are (1)
      background AGNs (group A non-thermal), (2) WD binaries (group A thermal), (3)
      X-ray active stars during the flare (group B) and (4) X-ray active stars in
      quiescence (group C).

\item We made a composite X-ray spectrum of each population and derived their fractional
      contributions to the GRXE separately for the hard-band continuum and the Fe K
      emission line feature. We found that the group A thermal and group B sources
      account for most of the Fe K emission.

\item We constructed the $\log N$--$\log S$ curves for each group in the flux range
      between $\approx$10$^{-13}$~$\ergcms$ and $\approx$10$^{-16}$~$\ergcms$ in the
      2--8~keV band.  We found that extra-Galactic sources dominate the flux range down
      to $\approx5\times10^{-15}$~$\ergcms$, which is taken over by the Galactic
      populations below this flux. For the Galactic sources, WD binaries such as
      magnetic and non-magnetic CVs, pre-CVs, and symbiotic stars, account for a larger
      fraction than X-ray active stars in the studied flux ranges.
\end{enumerate}

\acknowledgments
We acknowledge J. Hong, M. van den Berg and M. Sugizaki for useful discussion. We also
appreciate critical comments both by the reviewer and the editor. This research has made
use of public data obtained from \textit{Chandra} X-Ray Center which is operated for
NASA by the Smithsonian Astrophysical Observatory. This work has also made use of
software from High Energy Astrophysics Science Archive Research Center (HEASAC) which is
provided by NASA Goddard Space Flight Center. K.\,M. is financially supported by the
University of Tokyo global COE program and the Hayakawa Satio Foundation by the
Astronomical Society of Japan.

{\it Facilities:} \facility{CXO (ACIS)}.

\clearpage
\bibliographystyle{apj}

\bibliography{ms}

\end{document}

%% file: t01.tex
\begin{deluxetable}{cccccc}
\tabletypesize{\scriptsize}
\tablecolumns{11} 
\tablewidth{0pc} 
\tablecaption{Observation Log\label{t01}}
\tablehead{ 
\colhead{ObsID}  &  \colhead{Date} & \multicolumn{2}{c}{Coordinate (J2000.0)} &   \colhead{\textit{t}$_{\rm{exp}}$\tablenotemark{a}} & \colhead{Roll} \\
\cline{3-4}
\colhead{}   & \colhead{}      & \colhead{R.A.} & \colhead{Decl.}  & \colhead{(ks)}   & \colhead{(degree)}}
\startdata 
9500 & 2008-07-20 & 17:54:38.5 & $-$29:35:47 & 162.56 & 279.99\\
9501 & 2008-07-23 & 17:54:39.1 & $-$29:35:53 & 131.01 & 278.91\\
9502 & 2008-07-17 & 17:54:39.8 & $-$29:35:58 & 164.12 & 281.16\\
9503 & 2008-07-28 & 17:54:40.2 & $-$29:36:60 & 102.31 & 275.21\\
9504 & 2008-08-02 & 17:54:40.8 & $-$29:36:12 & 125.42 & 275.21\\
9505 & 2008-05-07 & 17:54:37.5 & $-$29:34:47 & \phantom{0}10.72 & \phantom{0}82.22\\
9854 & 2008-07-27 & 17:54:41.5 & $-$29:36:17 & \phantom{0}22.78 & 277.72\\
9855 & 2008-05-08 & 17:54:37.5 & $-$29:34:47 & \phantom{0}55.94 & \phantom{0}82.22\\
9892 & 2008-07-31 & 17:54:40.2 & $-$29:36:60 &  \phantom{0}65.79 & 275.21\\
9893 & 2008-08-01 & 17:54:40.8 & $-$29:36:12 & \phantom{0}42.16 & 275.21
\enddata 
\tablenotetext{a}{Exposure time.}
\end{deluxetable}

%% file: t02.tex
\setlength{\tabcolsep}{0.0008in}
\begin{deluxetable*}{rcccrrrrrrrrrccrrcrr}
\tighten
\centering
\tabletypesize{\tiny}
\tablecolumns{21} 
\tablewidth{0pt} 
\tablecaption{{\it Chandra} Catalog:  Basic Source Properties\label{t02}}
\tablehead{
  \multicolumn{2}{c}{Source} &
  \multicolumn{4}{c}{Position} &
  \multicolumn{5}{c}{Extracted Counts} &
  \multicolumn{9}{c}{Characteristics}  \\
\multicolumn{2}{c}{\hrulefill} &
\multicolumn{4}{c}{\hrulefill} &
\multicolumn{5}{c}{\hrulefill} &
\multicolumn{9}{c}{\hrulefill} \\
%
  \colhead{Seq}  &  \colhead{CXOU J} & 
  \colhead{$\rm{R.\,A.}$} & \colhead{$\rm{Decl.}$}
 &\colhead{Err} & \colhead{$\theta$}&
  \colhead{$C_{\rm{net}}$}& \colhead{$\Delta C_{\rm{net}}$}  & \colhead{$C^{\prime}_{\rm{bkg}}$} &
  \colhead{$C_{\rm{net,hard}}$} & \colhead{PSF} & \colhead{PS}&
  \colhead{$P_{\mathrm{B}}$}& \colhead{Anom} & \colhead{Var}& \colhead{EffExp} & \colhead{ME} & \colhead{Photo~\fx}  & \colhead{$q_{1}$}& \colhead{$q_{2}$}\\
  \colhead{\#}  &  \colhead{} & 
  \colhead{(deg)} & \colhead{(deg)} &  \colhead{(\arcsec)} & \colhead{(\arcmin)}&
  \colhead{}& \colhead{} & \colhead{} & \colhead{} & \colhead{Frac} & 
  \colhead{}& \colhead{}& \colhead{} & \colhead{}& \colhead{(ks)} &
  \colhead{(keV)}& \colhead{(ergs cm$^{-2}$ s$^{-1}$)}& \colhead{}& \colhead{}\\
  \colhead{(1)}  &  \colhead{(2)} 
& \colhead{(3)} & \colhead{(4)} &\colhead{(5)} 
& \colhead{(6)}& \colhead{(7)}  &  \colhead{(8)} & \colhead{(9)}&\colhead{(10)} 
& \colhead{(11)}& \colhead{(12)}  &  \colhead{(13)} & \colhead{(14)} &
 \colhead{(15)} & \colhead{(16)}& \colhead{(17)}& \colhead{(18)} & \colhead{(19)}&
 \colhead{(20)}
}
\startdata
1 & 175044.88$-$292837.6 &267.687010 & $-$29.477113 & 0.4 & 11.6 &  78.4 & 23.5 & 437.6 & 29.2 & 0.89 &  3.3 & $-$3.9 & .... & b & 486.5 & 1.5 & 1.6$\times$10$^{-15}$ & $-$0.80 & 0.63 \\
2 & 175045.10$-$293431.4 &267.687940 & $-$29.575409 & 0.4 & 9.2 &  38.4 & 14.9 & 163.6 & 30.0 & 0.90 &  2.5 & $-$2.7 & g... & ... & 325.1 & 3.0 & 1.1$\times$10$^{-15}$ & $-$0.81 & 1.26 \\
3 & 175045.10$-$293240.5 &267.687950 & $-$29.544603 & 0.3 & 9.7 &  56.1 & 19.1 & 279.9 & 17.7 & 0.89 &  2.9 & $-$3.3 & g... & ... & 535.5 & 1.5 & 2.4$\times$10$^{-15}$ & $-$0.31 & 1.63 \\
4 & 175045.51$-$293222.4 &267.689630 & $-$29.539560 & 0.3 & 9.7 &  46.4 & 16.5 & 201.6 & 34.1 & 0.78 &  2.7 & $-$3.1 & g... & ... & 597.9 & 2.9 & 1.9$\times$10$^{-15}$ & $-$0.33 & 1.57 \\
5 & 175045.86$-$293503.5 &267.691120 & $-$29.584322 & 0.3 & 9.1 &  60.6 & 16.5 & 190.4 & 52.7 & 0.90 &  3.6 & $-$5.0 & g... & ... & 417.5 & 2.7 & 3.0$\times$10$^{-15}$ & $-$0.39 & 1.67 \\
6 & 175045.90$-$293208.7 &267.691270 & $-$29.535760 & 0.3 & 9.7 &  92.1 & 20.3 & 291.9 & 24.2 & 0.85 &  4.4 & $<-$5 & g... & ... & 639.9 & 1.4 & 1.5$\times$10$^{-15}$ & $-$0.88 & 0.96 \\
7 & 175046.03$-$293539.1 &267.691800 & $-$29.594220 & 0.4 & 9.0 &  40.2 & 15.5 & 179.8 & 10.7 & 0.90 &  2.5 & $-$2.7 & g... & ... & 387.2 & 1.8 & 1.3$\times$10$^{-15}$ & $-$0.66 & 1.85 \\
8 & 175046.07$-$293405.8 &267.691960 & $-$29.568297 & 0.3 & 9.2 &  69.0 & 19.2 & 270.0 & 46.8 & 0.90 &  3.5 & $-$4.7 & g... & ... & 570.4 & 2.4 & 2.0$\times$10$^{-15}$ & $-$0.46 & 1.40 \\
9 & 175046.25$-$292947.1 &267.692710 & $-$29.496437 & 0.3 & 10.7 &  60.1 & 17.6 & 224.9 & 51.2 & 0.69 &  3.3 & $-$4.3 & .... & b & 678.2 & 3.3 & 2.5$\times$10$^{-15}$ & $-$0.22 & 1.35 \\
10 & 175046.62$-$292935.0 &267.694250 & $-$29.493066 & 0.3 & 10.8 &  58.6 & 17.7 & 229.4 & 73.7 & 0.70 &  3.2 & $-$4.1 & .... & b & 677.5 & 5.2 & 3.9$\times$10$^{-15}$ & \phantom{0}0.23 & 2.01 \\
11 & 175047.08$-$293334.4 &267.696170 & $-$29.559558 & 0.3 & 9.1 &  48.4 & 19.9 & 316.6 & 8.2 & 0.90 &  2.4 & $-$2.4 & g... & ... & 679.6 & 1.2 & 6.3$\times$10$^{-16}$ & $-$0.96 & 1.12 \\
12 & 175047.51$-$293733.0 &267.697980 & $-$29.625834 & 0.3 & 9.0 &  51.4 & 17.4 & 226.6 & 42.1 & 0.90 &  2.9 & $-$3.3 & g... & ... & 519.2 & 2.8 & 2.0$\times$10$^{-15}$ & $-$0.37 & 1.35 \\
13 & 175048.28$-$293250.7 &267.701180 & $-$29.547440 & 0.2 & 9.0 &  148.7 & 22.4 & 321.3 & 125.7 & 0.90 &  6.5 & $<-$5 & .... & a & 714.8 & 3.5 & 4.8$\times$10$^{-15}$ & $-$0.18 & 1.34 \\
14 & 175048.39$-$292816.1 &267.701660 & $-$29.471139 & 0.4 & 11.3 &  71.4 & 20.2 & 304.6 & 47.4 & 0.90 &  3.5 & $-$4.5 & g... & ... & 386.5 & 2.4 & 3.1$\times$10$^{-15}$ & $-$0.46 & 1.39 \\
15 & 175048.54$-$293937.5 &267.702280 & $-$29.660423 & 0.3 & 9.5 &  102.3 & 19.1 & 235.7 & 82.1 & 0.90 &  5.2 & $<-$5 & g... & ... & 457.0 & 3.0 & 5.1$\times$10$^{-15}$ & $-$0.29 & 1.45 \\
16 & 175048.65$-$293538.6 &267.702740 & $-$29.594083 & 0.2 & 8.5 &  127.2 & 18.6 & 194.8 & 79.1 & 0.80 &  6.6 & $<-$5 & g... & ... & 734.4 & 3.6 & 4.8$\times$10$^{-15}$ & $-$0.15 & 0.79 \\
17 & 175048.86$-$293736.9 &267.703620 & $-$29.626938 & 0.3 & 8.8 &  73.8 & 20.0 & 296.2 & 30.3 & 0.90 &  3.6 & $-$4.8 & g... & ... & 698.8 & 1.9 & 1.4$\times$10$^{-15}$ & $-$0.65 & 0.84 \\
18 & 175048.91$-$292906.9 &267.703810 & $-$29.485275 & 0.3 & 10.6 &  124.8 & 25.5 & 482.2 & 45.0 & 0.90 &  4.8 & $<-$5 & .... & a & 668.0 & 1.6 & 1.9$\times$10$^{-15}$ & $-$0.78 & 1.14 \\
19 & 175048.95$-$293547.9 &267.703980 & $-$29.596661 & 0.2 & 8.5 &  82.7 & 16.4 & 165.3 & 51.4 & 0.76 &  4.9 & $<-$5 & .... & a & 752.1 & 2.5 & 2.1$\times$10$^{-15}$ & $-$0.43 & 0.77 \\
20 & 175049.07$-$293803.6 &267.704500 & $-$29.634354 & 0.3 & 8.8 &  82.8 & 20.6 & 312.2 & 13.8 & 0.90 &  3.9 & $<-$5 & g... & ... & 702.2 & 1.6 & 1.3$\times$10$^{-15}$ & $-$0.79 & 1.45
\enddata
\tablecomments{
{\bf Col.(1):} X-ray catalog sequence number, sorted by R.\,A.
{\bf Col.(2):} IAU designation.
{\bf Col.(3),(4):} R.\,A. and Decl. in the equinox J2000.0.
{\bf Col.(5):} Estimated standard deviation of the position error.
{\bf Col.(6):} Off-axis angle.
{\bf Col.(7),(8):} Estimated net counts and errors extracted in 0.5--8.0~keV.
{\bf Col.(9):} Background counts in 0.5--8~keV.
{\bf Col.(10):} Estimated net counts extracted in the hard band (2.0--8.0~keV).
{\bf Col.(11):} Fraction of the PSF (at 1.497 keV) enclosed within the extraction
region. A reduced PSF fraction (significantly below 90\%) may indicate that
the source is in a crowded region.
{\bf Col.(12):} Photometric significance.
{\bf Col.(13):} Logarithmic probability that the extracted counts (0.5--8.0~keV) are solely
from background. 
{\bf Col.(14):} Source anomalies: g = fractional time that source was on a detector
(FRACEXPO from {\em mkarf}) is $<0.9$ 
{\bf Col.(15):} Variability characterization based on K-S statistic (0.5--8.0~keV): a =
no evidence for variability ($0.05<P_{KS}$); b = possibly variable
($0.005<P_{KS}<0.05$); c = definitely variable ($P_{KS}<0.005$).  No value is reported
for sources in chip gaps or on field edges.
{\bf Col.(16):} Effective exposure time: approximate time the source would have to be
observed on-axis to obtain the reported number of counts.
{\bf Col.(17):} Background-corrected median photon energy (0.5--8.0~keV).
{\bf Col.(18):} Photometric flux estimate in the 0.5--8.0~keV band. 
Table~\ref{t02} is published in its entirety in the electronic edition of the {\it
 Astrophysical Journal}.  A portion is shown here for guidance regarding its form and
 content.
{\bf Col.(19), (20):} $q_{1}$ and $q_{2}$ quantile values indicating the spectral shape
 of each source (see text in \S~\ref{s4-1-1}).
}
%
\end{deluxetable*}

%% file: t03a.tex
\setlength{\tabcolsep}{0.05in}
\begin{deluxetable*}{rcrrlllrrrrrr}
\tabletypesize{\scriptsize}
\tablecolumns{13}
\tablewidth{0pt}
\tablecaption{Thermal fittings for the X-ray sources with source counts over 1000\label{t03a}}
\tablehead{
 \multicolumn{4}{c}{Source\tablenotemark{1}} &
 \multicolumn{3}{c}{Spectral Fit\tablenotemark{2}} &     
 \multicolumn{5}{c}{X-ray Flux\tablenotemark{3}} & 
 \multicolumn{1}{c}{} \\
\multicolumn{4}{c}{\hrulefill} &
\multicolumn{3}{c}{\hrulefill} &
\multicolumn{5}{c}{\hrulefill} & 
\multicolumn{1}{c}{\hrulefill} \\
\multicolumn{1}{c}{Seq.} & 
\multicolumn{1}{c}{CXOU J} & 
\multicolumn{1}{c}{$C_{\mathrm{t,net}}$} & 
\multicolumn{1}{c}{Signif.} &
\multicolumn{1}{c}{$\log N_{\mathrm{H}}$} & 
\multicolumn{1}{c}{$kT_{1}$} & 
\multicolumn{1}{c}{$kT_{2}$} &  
\multicolumn{1}{c}{$F_{\mathrm{s}}$} & 
\multicolumn{1}{c}{$F_{\mathrm{h}}$} & 
\multicolumn{1}{c}{$F_{\mathrm{h,c}}$} & 
\multicolumn{1}{c}{$F_{\mathrm{t}}$} & 
\multicolumn{1}{c}{$F_{\mathrm{t,c}}$} & 
\multicolumn{1}{c}{$\chi^{2}$/d.o.f.}  \\
\multicolumn{1}{c}{\#} & 
\multicolumn{1}{c}{} & 
\multicolumn{1}{c}{} & 
\multicolumn{1}{c}{} &
\multicolumn{1}{c}{(cm$^{-2}$)} & 
\multicolumn{1}{c}{(keV)} & 
\multicolumn{1}{c}{(keV)} & 
\multicolumn{5}{c}{(ergs cm$^{-2}$ s$^{-1}$)} & 
\multicolumn{1}{c}{} \\
\multicolumn{1}{c}{(1)} & 
\multicolumn{1}{c}{(2)} & 
\multicolumn{1}{c}{(3)} & 
\multicolumn{1}{c}{(4)} &
\multicolumn{1}{c}{(5)} & 
\multicolumn{1}{c}{(6)} & 
\multicolumn{1}{c}{(7)} & 
\multicolumn{1}{c}{(8)} &
\multicolumn{1}{c}{(9)} &
\multicolumn{1}{c}{(10)} &
\multicolumn{1}{c}{(11)} &
\multicolumn{1}{c}{(12)} & 
\multicolumn{1}{c}{(13)} 
}
\startdata
287 & 175107.66$-$294037.3 & 3252.3 & 55.1 &$21.6_{-0.07}^{+0.07}$ & $0.3_{-0.1}^{+0.1}$ & $3.9_{-0.5}^{+0.7}$ & 3.6e-14 & 2.6e-14  &2.6e-14  & 5.0e-14 & 5.2e-14 & 129.30/131\\
1775 & 175148.91$-$293505.6 & 1779.1 & 40.7 &$21.6_{-0.07}^{+0.07}$ &
 $1.5_{-0.1}^{+0.1}$ & ... &1.1e-14 & 8.1e-15 & 8.6e-15 & 1.9e-14 &
 3.0e-14 &103.52/\phantom{0}78
\enddata
\tablecomments{
{\bf 1:} For convenience, cols.\ (1)--(4) reproduce the source identification, net
 counts, and photometric significance data from Table~\ref{t01}, which are sorted by the
 number of counts.
{\bf 2:} All fits used the {\em source} model ``tbabs*vapec'' in XSPEC abundances frozen at the values relative to  
\citet{Anders89}, scaled to \citet{Wilms00}, using the tbabs absorption code in XSPEC.  
Cols.\ (5), (6), and (7) present the best-fit values for the extinction column density and plasma temperature parameters.
Uncertainties represent 90\% confidence intervals.
More significant digits are used for uncertainties $<$0.1 in order to avoid large rounding errors; for consistency, the same number of significant digits is used for both lower and upper uncertainties.
Uncertainties are missing when XSPEC was unable to compute them or when their values were so large that the parameter is effectively unconstrained.  
Fits lacking uncertainties, fits with large uncertainties, and fits with frozen parameters should be viewed merely as splines to the data to obtain rough estimates of luminosities; the listed parameter values are not robust.
{\bf 3:} X-ray flux derived from the model spectrum are presented in cols.\ (8)--(12): 
(s) soft band (0.5--2 keV); (h) hard band (2--8 keV); (t) total band (0.5--8 keV).
Absorption-corrected fluxes are subscripted with a $c$.
Cols. (8) and (12) are omitted when $\log N_{\mathrm{H}} > 22.5$~cm$^{-2}$ since the soft band emission 
is essentially unmeasurable.
}
\end{deluxetable*}

%% file: t03b.tex
\setlength{\tabcolsep}{0.05in}
\begin{deluxetable*}{rcrrlllrrrrrr}
\tighten
\tabletypesize{\scriptsize}
\tablecolumns{13}
\tablewidth{0pt}
\tablecaption{Power-law fitting for the X-ray sources with source counts over 1000\label{t03b}}
\tablehead{
 \multicolumn{4}{c}{Source\tablenotemark{1}} &
 \multicolumn{3}{c}{Spectral Fit\tablenotemark{2}} &
 \multicolumn{5}{c}{X-ray Flux\tablenotemark{3}} &
 \multicolumn{1}{c}{} \\
\multicolumn{4}{c}{\hrulefill} &
\multicolumn{3}{c}{\hrulefill} &
\multicolumn{5}{c}{\hrulefill} &
\multicolumn{1}{c}{\hrulefill} \\
\multicolumn{1}{c}{Seq.} & 
\multicolumn{1}{c}{CXOU J} & 
\multicolumn{1}{c}{$C_{\mathrm{t,net}}$} & 
\multicolumn{1}{c}{Signif.} &
\multicolumn{1}{c}{$\log N_{\mathrm{H}}$} & 
\multicolumn{1}{c}{$\Gamma$} & 
\multicolumn{1}{c}{$\log N_{\Gamma}$} &  
\multicolumn{1}{c}{$F_{\mathrm{s}}$}&
\multicolumn{1}{c}{$F_{\mathrm{h}}$} & 
\multicolumn{1}{c}{$F_{\mathrm{h,c}}$} & 
\multicolumn{1}{c}{$F_{\mathrm{t}}$} & 
\multicolumn{1}{c}{$F_{\mathrm{t,c}}$} &
\multicolumn{1}{c}{$\chi^{2}$/d.o.f.} \\
\multicolumn{1}{c}{\#} & 
\multicolumn{1}{c}{} & 
\multicolumn{1}{c}{} & 
\multicolumn{1}{c}{} &
\multicolumn{1}{c}{(cm$^{-2}$)} & 
\multicolumn{1}{c}{} & 
\multicolumn{1}{c}{} & 
\multicolumn{5}{c}{(ergs cm$^{-2}$ s$^{-1}$)} &
\multicolumn{1}{c}{}\\
\multicolumn{1}{c}{(1)} & 
\multicolumn{1}{c}{(2)} & 
\multicolumn{1}{c}{(3)} & 
\multicolumn{1}{c}{(4)} &
\multicolumn{1}{c}{(5)} & 
\multicolumn{1}{c}{(6)} & 
\multicolumn{1}{c}{(7)} & 
\multicolumn{1}{c}{(8)} &
\multicolumn{1}{c}{(9)} &
\multicolumn{1}{c}{(10)} &
\multicolumn{1}{c}{(11)} &
\multicolumn{1}{c}{(12)} &
\multicolumn{1}{c}{(13)}}
\startdata
1895 & 175154.71-292806.1  & 6440.6 & 77.1 &$22.0_{-0.01}^{+0.01}$ & 1.8$_{-0.1}^{+0.1}$ &  $-4.26_{-0.06}^{+0.06}$ & 1.7e-14 & 1.4e-13 & 1.7e-13 & 1.6e-13 & 
2.9e-13 & 300.71/244\\
438 & 175113.70-293110.1 & 4325.4 & 64.4 &$22.5_{-0.04}^{+0.04}$ & 1.5$_{-0.1}^{+0.1}$ &
 $-4.53_{-0.08}^{+0.08}$ &5.8e-15 & 1.1e-13 & 1.3e-13 & 1.2e-13 & 2.0e-13 & 196.80/203\\
1149 & 175131.68-292957.0 & 2908.3 & 52.5 &$22.4_{-0.06}^{+0.06}$ & 1.3$_{-0.1}^{+0.1}$
 & $-4.87_{-0.09}^{+0.10}$ & 4.7e-15 & 7.8e-14 & 8.7e-14 & 8.2e-14 & 1.2e-13 &
 137.31/142\\
66 & 175054.26-294327.4 & 1600.2      & 46.2    & $22.7_{-0.05}^{+0.04}$     & 1.5$_{-0.2}^{+0.2}$
 & $-4.87_{-0.09}^{+0.10}$ &7.0e-16   & 8.0e-14 &9.4e-14  &8.0e-14  &9.6e-14  &
 128.96/110\\
33 & 175051.16-293419.5 & 1469.3 & 34.5 &$22.4_{-0.09}^{+0.09}$ & 1.7$_{-0.2}^{+0.3}$ &
 $-4.92_{-0.2}^{+0.2}$  & 3.7e-15 & 3.7e-14 & 4.0e-14 & 4.1e-14 & 5.7e-14 & 88.60/\phantom{0}71\\
1831 & 175151.29-293310.3 & 1447.3 & 36.3 &$22.3_{-0.10}^{+0.09}$ & $0.65_{-0.2}^{+0.2}$
  & $-5.51_{-0.1}^{+0.1}$ & 1.8e-15 & 4.8e-14 & 5.2e-14 & 5.0e-14 & 5.8e-14 & 113.47/\phantom{0}78\\
1252 & 175134.06-293103.9 & 1390.3 & 36.0 &$22.1_{-0.08}^{+0.08}$ & 1.6$_{-0.2}^{+0.2}$
 & $-5.18_{-0.09}^{+0.10}$ &3.9e-15 & 2.5e-14 & 2.6e-14 & 2.9e-14 & 4.1e-14 & 75.74/\phantom{0}74\\
1869 & 175153.33-294245.0 & 1202.7 & 30.0 &$22.4_{-0.1}^{+0.1}$ & 1.0$_{-0.3}^{+0.3}$ &
  $-5.30_{-0.2}^{+0.2}$ & 3.3e-15 & 5.0e-14 & 5.1e-14 & 5.3e-14 & 5.8e-14 &
 60.59/\phantom{0}58
\enddata
\tablecomments{
{\bf 1:} For convenience, cols.\ (1)--(4) reproduce the source identification, net
 counts, and photometric significance data from Table~\ref{t01}, which are sorted by the number of counts.
{\bf 2:} All fits used the {\em source} model ``tbabs*powerlaw'' in XSPEC.  
Cols.\ (5) and (6) present the best-fit values for the extinction column density and power law photon index parameters.
Col.\ (7) presents the power law normalization for the model spectrum. 
Uncertainties represent 90\% confidence intervals.
More significant digits are used for uncertainties $<$0.1 in order to avoid large
 rounding errors; for consistency, 
the same number of significant digits is used for both lower and upper uncertainties.
Uncertainties are missing when XSPEC was unable to compute them or when their values were so large that the parameter is effectively unconstrained.  
Fits lacking uncertainties, fits with large uncertainties, and fits with frozen parameters should be viewed merely as splines to the data to obtain rough estimates of luminosities; the listed parameter values are unreliable.
{\bf 3:} X-ray Flux derived from the model spectrum, assuming a distance
 of 8.5~kpc, are presented in cols.\ (8)--(12): (s) = soft band (0.5--2 keV); (h) hard
 band (2--8 keV); (t) total band (0.5--8 keV). Absorption-corrected fluxes are subscripted with a $c$.
Cols. (8) and (12) are omitted when $\log N_{\mathrm{H}} > 22.5$~cm$^{-2}$ since the soft band emission 
is essentially unmeasurable.
}
\end{deluxetable*}

%% file: t05.tex
\begin{deluxetable*}{llccccccc}
 \tabletypesize{\scriptsize}
 \tablecolumns{9} 
 \tablewidth{0pc} 
 \tablecaption{Best-fit parameters of the power-law $+$ Gaussian model for the Fe K feature\tablenotemark{a}\label{t05}}
 \tablehead{
 \colhead{Group} & 
 \colhead{$E_{\mathrm{gau}}$\tablenotemark{b}} &
 \colhead{EW$_{\mathrm{gau}}$\tablenotemark{c}} & 
 \colhead{$F_\mathrm{pow}$/$10^{-13}$\tablenotemark{d}} &
 \colhead{$f_\mathrm{pow}^{\mathrm{(PS)}}$\tablenotemark{e}} & 
 \colhead{$f_\mathrm{pow}^{\mathrm{(Total)}}$\tablenotemark{f}} & 
 \colhead{$F_\mathrm{gau}$/$10^{-13}$\tablenotemark{g}} &
 \colhead{$f_\mathrm{gau}^{\mathrm{(PS)}}$\tablenotemark{h}} & 
 \colhead{$f_\mathrm{gau}^{\mathrm{(Total)}}$\tablenotemark{i}} \\
 & 
 \colhead{(keV)} &
 \colhead{(eV)} &
 \colhead{($\ergcms$)} &
&
&
 \colhead{($\ergcms$)} &
&
}
\startdata
 A & 6.69$_{-0.05}^{+0.06}$ & 268$_{-99}^{+244}$ & \phantom{0}12.02$_{-0.54}^{+0.58}$ & 0.68$_{-0.03}^{+0.04}$ & 0.44$_{-0.02}^{+0.02}$ &0.98$_{-0.12}^{+0.12}$ & 0.62$_{-0.08}^{+0.08}$ & 0.26$_{-0.03}^{+0.03}$\\
 B & 6.70$_{-0.08}^{+0.08}$&  413$_{-216}^{+263}$ & \phantom{0}5.01$_{-0.55}^{+0.61}$ & 0.28$_{-0.03}^{+0.04}$ & 0.18$_{-0.02}^{+0.02}$ &0.50$_{-0.19}^{+0.34}$ & 0.32$_{-0.12}^{+0.22}$ & 0.13$_{-0.05}^{+0.09}$\\
 C & \nodata                & \nodata             & \phantom{0}0.68$_{-0.12}^{+0.19}$ & 0.04$_{-0.01}^{+0.01}$ & 0.02$_{-0.01}^{+0.01}$ & $<$0.04  & \nodata & \nodata\\
\cline{1-9}
 All PS& 6.69$_{-0.04}^{+0.05}$   &  321$_{-122}^{+116}$ & 17.18$_{-0.70}^{+0.71}$ &\nodata & \nodata& 1.58$_{-0.51}^{+0.60}$ &\nodata  & \nodata\\
 Total emission & 6.74$_{-0.04}^{+0.05}$   &  581$_{-131}^{+100}$ &
 27.22$_{-0.54}^{+0.54}$ &\nodata  & \nodata & 3.82$_{-0.51}^{+0.50}$ & \nodata & \nodata
\enddata
\tablenotetext{a}{The best-fit value and a 1$\sigma$ statistical uncertainty are
 given. For those with no best-fit values, only the 1 $\sigma$ upper limit is given.}
\tablenotetext{b}{Center energy of the Gaussian component.}
\tablenotetext{c}{Equivalent width of the Gaussian component.}
\tablenotetext{d}{Flux of the power-law component in the 4--8~keV band.}
\tablenotetext{e}{Fractional flux of the power-law component among all point sources.}
\tablenotetext{f}{Fractional flux of the power-law component among total GRXE spectrum.}
\tablenotetext{g}{Flux of the Gaussian component.}
\tablenotetext{h}{Fractional flux of the Gaussian component among all point sources.}
\tablenotetext{i}{Fraction flux of the Gaussian component among total GRXE spectrum.}
\end{deluxetable*}

%% file: t06.tex
\begin{deluxetable*}{ccccccccc}
\tabletypesize{\scriptsize}
\tablecolumns{9} 
\tablewidth{0pc} 
\tablecaption{Best-fit parameters for global spectral model in 0.5--8.0~keV\tablenotemark{a}\label{t06}}
\tablehead{ 
\colhead{Group}  &  
\colhead{\nh$^{(1)}$\tablenotemark{b}} & 
\colhead{\kt$^{(1)}$\tablenotemark{c}} &
\colhead{\nh $^{(2)}$\tablenotemark{d}} & 
\colhead{\kt$^{(2)}$\tablenotemark{e}} &
\colhead{Z\tablenotemark{f}} &
\colhead{$\Gamma$\tablenotemark{g}} &
\colhead{$N_{\mathrm{apec}}^{(1)}$/$N_{\mathrm{apec}}^{(2)}$\tablenotemark{h}} &
\colhead{$\chi^{2}$/d.o.f.\tablenotemark{i}} \\
\colhead{}  &  
\colhead{(10$^{22}$~cm$^{-2}$)} & 
\colhead{(keV)} & 
\colhead{(10$^{22}$~cm$^{-2}$)} & 
\colhead{(keV)} &
\colhead{} &
\colhead{} &
\colhead{} &
\colhead{} 
}
\startdata 
A         & 1.09$_{-0.50}^{+0.39}$ & 6.65$_{-3.03}^{+3.24}$ & 2.46$_{-0.58}^{+2.35}$ &
 \nodata &  0.97$_{-0.32}^{+0.36}$ &1.29$_{-0.40}^{+0.18}$ & \nodata  &\phantom{0}205.36/504\\
B         & 0.75$_{-0.05}^{+0.06}$ & 0.74$_{-0.45}^{+0.54}$ & 0.80$_{-0.18}^{+0.22}$ &
 7.87$_{-4.84}^{+1.86}$ & 0.99$_{-0.29}^{+0.33}$ & \nodata & 0.47$_{-0.14}^{+0.18}$ &\phantom{0}98.68/103\\
C        & 0.70$_{-0.11}^{+0.18}$  & 0.78$_{-0.03}^{+0.04}$ & 0.04$_{-0.04}^{+0.05}$ &
 4.50$_{-0.35}^{+0.65}$ & 0.15$_{-0.12}^{+0.16}$ & \nodata & 6.70$_{-0.20}^{+0.13}$ &\phantom{0}85.86/102
\enddata
\tablenotetext{a}{The best-fit value and a 1$\sigma$ statistical uncertainty are given.}
\tablenotetext{b}{Interstellar extinction column density for the first component (the
 lower temperature component for the two-temperature model).}
 \tablenotetext{c}{Plasma temperature for the first component (the
 lower temperature component for the two-temperature model).}
\tablenotetext{d}{Interstellar extinction column density for the second component (the
 higher temperature component for the two-temperature model).}
 \tablenotetext{e}{Plasma temperature for the second component (the
 higher temperature component for the two-temperature model).}
 \tablenotetext{f}{Metal abundance relative to the solar value for the thermal component.}
 \tablenotetext{g}{Photon index for the power-law model as the second component.}
 \tablenotetext{h}{Normalization ratio between the first (the
 lower temperature) and and the second (the higher temperature) component for the
 two-temperature model.}
 \tablenotetext{i}{The goodness of the fit.}
\end{deluxetable*}